\documentclass[review]{elsarticle}

\usepackage{amsthm}
\usepackage{subfig}
\usepackage[table,x11names]{xcolor}
\usepackage{amssymb}
\usepackage{amsmath}
\newtheorem{mydef}{Definition}

\usepackage{smartdiagram}
\usepackage{graphicx}
\usepackage{soul}
\usepackage{float}
\usepackage{multirow}
\usepackage{booktabs} 
\usepackage{etoolbox} 
\usepackage[vlined,boxed,commentsnumbered, ruled, linesnumbered]{algorithm2e}
\SetKw{KwBy}{by}
\usepackage{tablefootnote}
\makeatletter
\newcommand{\removelatexerror}{\let\@latex@error\@gobble}
\makeatother

\def\checkmark{\tikz\fill[scale=0.4](0,.35) -- (.25,0) -- (1,.7) -- (.25,.15) -- cycle;}

\usepackage{smartdiagram}
\usepackage[margin=3cm]{geometry}

\usepackage{footnote}

\usepackage[margin=3cm]{geometry}

\makeatletter
\def\BState{\State\hskip-\ALG@thistlm}
\makeatother

\usepackage{setspace}

\usepackage{breakcites}
\usepackage{enumerate}
\usepackage{makecell}

\usepackage{float}
\usepackage{listings}

\usepackage{tikz}
\usetikzlibrary{trees}
\usepackage{paralist}
\usepackage{ragged2e}
\usepackage[framemethod=tikz]{mdframed}

\usetikzlibrary{positioning,arrows.meta}

\definecolor{arrowblue}{RGB}{98,145,224}

\usepackage{soul}

\usepackage[colorinlistoftodos]{todonotes}

\usepackage{lineno,hyperref}
\usepackage{cleveref}
\usepackage{setspace}
\modulolinenumbers[1]

\journal{Journal of Computers  $\&$ Security}

\begin{document}

\begin{frontmatter}

\title{Privacy Preserving Face Recognition Utilizing Differential Privacy}

\author[mymainaddress,mysecondaryaddress]{M.A.P.~Chamikara
	\corref{mycorrespondingauthor}}
\cortext[mycorrespondingauthor]{Corresponding author}
\ead{pathumchamikara.mahawagaarachchige@rmit.edu.au}

\author[mymainaddress]{P.~Bertok}
\author[mymainaddress]{I.~Khalil}
\author[mysecondaryaddress]{D.~Liu}
\author[mysecondaryaddress]{S.~Camtepe}


\address[mymainaddress]{RMIT University, Australia}
\address[mysecondaryaddress]{CSIRO Data61, Australia}

\begin{abstract}
\begin{mdframed}[backgroundcolor=green!50,rightline=false,leftline=false]
\centering 
The published article can be found at \url{https://doi.org/10.1016/j.cose.2020.101951}
\end{mdframed}

Facial recognition technologies are implemented in many areas, including but not limited to, citizen surveillance, crime control, activity monitoring, and facial expression evaluation. However, processing biometric information is a resource-intensive task that often involves third-party servers, which can be accessed by adversaries with malicious intent.  Biometric information delivered to untrusted third-party servers in an uncontrolled manner can be considered a significant privacy leak (i.e. uncontrolled information release)  as biometrics can be correlated with sensitive data such as healthcare or financial records. In this paper, we propose a privacy-preserving technique for ``controlled information release'', where we disguise an original face image and prevent leakage of the biometric features while identifying a person. We introduce a new privacy-preserving face recognition protocol named PEEP (\underline{P}rivacy using \underline{E}ig\underline{E}nface \underline{P}erturbation) that utilizes local differential privacy. PEEP applies perturbation to Eigenfaces utilizing differential privacy and stores only the perturbed data in the third-party servers to run a standard Eigenface recognition algorithm. As a result, the trained model will not be vulnerable to privacy attacks such as membership inference and model memorization attacks. Our experiments show that PEEP exhibits a classification accuracy of around 70\% - 90\% under standard privacy settings.

\end{abstract}


\end{frontmatter}


\section{Introduction}
\label{intro}
Face recognition has many applications in the fields of image processing and computer vision; advancements in related technologies allow its efficient and accurate integration in many areas from individual face recognition for unlocking a mobile device to crowd surveillance.  Companies have also invested heavily in this field; Google's facial recognition in the Google Glass project~\cite{mandal2014wearable}, Facebook's DeepFace technology~\cite{macaulay2016queen}, and Apple's patented face identification system~\cite{bhagavatula2015biometric} are examples of the growing number of facial identification systems.  Existing face recognition technologies and the widespread use of biometrics introduce a serious threat to individuals' privacy, exacerbated by the fact that biometric identification is often done quietly, without proper consent from observed people. For example, the UK uses an estimated 4.2 million surveillance cameras to monitor public areas~\cite{erkin2009privacy}. However, it is not feasible to obtain explicit consent from an extremely large number of persons being watched. Nevertheless, facial images directly reflect the owners' identity, and they can be easily linked to other sensitive information such as health records and financial records, raising privacy concerns. Biometric data analysis systems often need to employ high-performance third-party servers to conduct complex computational operations on large numbers of biometric data inputs. However, these third-party servers can be accessed by untrusted parties causing privacy issues.

Among different definitions, information privacy can be defined as the ``controlled information release" that permits an anticipated level of utility via a private function that protects the identity of the data owners~\cite{chamikara2019efficient}. Privacy-preserving face recognition involves at least two main parties: one needs to recognize an image (party 1), and the other holds the database of images (party 2). Data encryption would allow party 1 to learn the result without learning the execution of the recognition algorithm or its parameters, whereas party 2 would not learn the input image or the result of the recognition process~\cite{erkin2009privacy}. However, the high computational complexity and the need to trust the parties for their respective responsibilities can be major issues. Proposed in this paper is data perturbation, which is significantly less computationally complex, but incurs a certain level of utility loss. Data perturbation allows all parties to be untrusted~\cite{chamikaraprocal}. The parties will learn only the classification result (e.g. name/tag of the image) with a certain level of confidence, but will not have access to the original image.  The literature identifies two major application scenarios of recognition technologies in which a third party server is used. They are (1) the use of biometric data such as face images and fingerprint to identify and authenticate a person (e.g. at border crossings) and (2) deploy surveillance cameras in public places to automatically match or identify faces (offender tracking/criminal investigations ~\cite{chamikara2016fuzzy}). There are a few methods that are based on encryption to provide privacy-preserving face recognition~\cite{erkin2009privacy,sadeghi2009efficient,xiang2016privacy}, which need one or more trusted third parties in a server-based setting (e.g. cloud servers).  However, in an environment where no trusted party is present, such semi-honest approaches raise privacy concerns, as the authorized trusted parties are still allowed to access the original image data (raw or encrypted). Moreover, an encryption-based mechanism for scenarios that process millions of faces would be extremely inefficient and difficult to maintain.  The methods such as $k-same$ ~\cite{newton2005preserving} for preserving privacy by de-identifying face images can avoid the necessity of a trusted third-party. However, such methods introduce utility issues in large scale scenarios with millions of faces, due to the limitations of the underlying privacy models used (e.g. $k-anonymity$)~\cite{chamikaraprocal}. We identify five main types of issues (TYIS) with the existing privacy-preserving approaches for face recognition. They are as follows. TYIS 1: face biometrics should not be linkable to other sensitive data, TYIS 2: the method should be scalable and resource friendly, TYIS 3: face biometrics should not be accessible by anyone (i.e. use one-way transformation), TYIS 4: face biometrics of the same person from two different applications should not be linkable, and TYIS 5: face biometrics should be revocable (if data is leaked, the application should have a way of revoking them to prevent any malicious use).

This paper proposes a method to control privacy leakage from face recognition, answering the five TYIS better than the existing privacy-preserving face recognition approaches. We propose an approach that stores data in a perturbed form. The method utilizes differential privacy to devise a novel technique (named PEEP: \underline{P}rivacy using \underline{E}ig\underline{E}nface \underline{P}erturbation) for privacy-preserving face recognition.  PEEP uses the properties of local differential privacy to apply perturbation on input image data to limit potential privacy leaks due to the involvement of untrusted third-party servers and users.  To avoid the necessity of a trusted third party, we apply randomization to the data used for training and testing. Due to the extremely low complexity, PEEP can be easily implemented on resource-constrained devices, allowing the possibility of perturbation at the input end. The ability to control the level of privacy via adjusting the privacy budget is an additional advantage of the proposed method. The privacy budget is used to signify the level of privacy provided by a privacy-preserving algorithm; the higher the privacy budget, the lower the privacy.  PEEP utilizes local differential privacy at the cost of as low as $6$ percent drop in accuracy (e.g. $85\%$ to $79\%$) with a privacy budget of $\varepsilon=8$.  A mechanism with a privacy budget ($\varepsilon$) of $0<\varepsilon\leq 9$ is considered to provide an acceptable level of privacy~\cite{abadi2016deep,arachchige2019local}. Consequently, PEEP is capable of adjusting the privacy-accuracy trade-off by changing the privacy budget through added noise.

The rest of the paper is organized as follows. Section \ref{relwork} provides a summary of existing related work. The foundations of the proposed work are briefly discussed in Section \ref{fndatins}. Section \ref{ourapprch} provides the technical details of the proposed approach. The results are discussed in Section \ref{resdis}. The paper is concluded in Section \ref{concls}. 

\section{Related Work}
\label{relwork}
Literature shows a vast advancement in the area of face recognition that has employed different approaches, such as input image preprocessing~\cite{heseltine2003face}, statistical approaches~\cite{tsalakanidou2003use,delac2005appearance}, and deep learning~\cite{parkhi2015deep}. The continuous improvements in the field have significantly improved the accuracy of face recognition making it a vastly used approach in many fields~\cite{parkhi2015deep}. Furthermore, the approaches, such as proposed by Cendrillon et al., show the dynamic capabilities of face recognition approaches that allow real-time processing~\cite{cendrillon2000real}. However, biometric data analysis is a vast area not limited to face recognition. With biometric data, a major threat is privacy violation~\cite{bhargav2007privacy}. Biometric data are almost always non-revocable and can be used to identify a person in a large set of individuals easily; hence, it is essential to apply some privacy-preserving mechanism when using biometrics, e.g. for identification and authentication~\cite{bringer2013privacy}. Literature shows a few approaches to address privacy issues in face recognition.  Zekeriya Erkin et al. (ZEYN)~\cite{erkin2009privacy} introduced a privacy-preserving face recognition method based on a cryptographic protocol for comparing two Pailler-encrypted values. Their solution focuses on a two-party scenario where one party holds the privacy-preserving algorithm and the database of face images, and the other party wants to recognize/classify a facial image input. ZEYN requires O(log M) rounds, and it needs computationally expensive operations on homomorphically encrypted data to recognize a face in a database of images, hence not suitable for large scale scenarios. Ahman-Reza Sadehi et al. (ANRA) ~\cite{sadeghi2009efficient} introduced a relatively efficient method based on homomorphic encryption with garbled circuits. Nevertheless, the complexity of ANRA also has the same problem of failing to address large scale scenarios. Xiang et al. tried to overcome the computational complexities of the previous methods by introducing another cryptographic mechanism that uses the cloud~\cite{xiang2016privacy} for outsourced computations. However, being a semi-honest model, introducing another untrusted module such as the cloud increases the possibility of privacy leak. PE-MIU (Privacy-Enhancing face recognition approach based on Minimum Information Units)~\cite{terhorst2020pe} and POR (lightweight privacy-preserving adaptive boosting (AdaBoost) classification framework for face recognition)~\cite{ma2019lightweight}  are two other recently developed privacy-preserving face recognition approaches. PE-MIU is based on the concept of minimum information units, whereas POR is based on additive secret sharing. PE-MIU is also a semi-honest approach, which lacks a proper privacy definition in its mechanism.  Moreover, the scalability of PE-MIU can be limited due to the exponential template comparisons necessary during the execution of the proposed algorithm. POR provides a relatively efficient approach compared to the previous encryption-based approaches. However, being a semi-honest approach, POR inherits the issues of any semi-honest approach discussed above. The proposed cryptographic methods cannot work without a trusted third party, and these trusted parties may later behave maliciously. Newton et al. proposed a de-identification approach for face images (named as $k-same$), which does not need complex cryptographic operations~\cite{newton2005preserving}. The proposed method is based on $k-anonymity$~\cite{chamikaraprocal,chamikara2019infosci}.  However, $k-anonymity$ tends to reduce accuracy and increase information leak when introduced with high dimensional data~\cite{chamikaraprocal}. The same problem can occur when using $k-same$ for large scale scenarios involving the surveillance of millions of people.  In addition to these works, researchers have looked at complementary techniques such as developing privacy-friendly surveillance cameras~\cite{dufaux2006scrambling,yu2008privacy}, but these methods do not provide sufficient accuracy for privacy-preserving face recognition.

Fingerprint data and iris data are two other heavily used biometrics for identification and authentication. Privacy-preserving finger code authentication~\cite{barni2010privacy}, and privacy-preserving key generation for iris biometrics~\cite{rathgeb2010privacy} are two approaches that apply cryptographic methods to maintain the privacy of fingerprint and iris data. However, these solutions also need more efficient procedures, as cryptographic approaches are inefficient in calculations~\cite{rathgeb2010privacy, gai2016privacy}. Privacy-preserving fingerprint and iris analysis can be possible future applications for PEEP, but this needs further investigation.  Classification is the most commonly applied data mining technique that is used in biometric systems~\cite{brady1999biometric}. Encryption and data perturbation are two main approaches also used for privacy-preserving data mining (PPDM)~\cite{ yang2017efficient}. Data perturbation often entails lower computational complexity than encryption at the expense of utility. Hence, data perturbation is better at producing high efficiency in large scale data mining.   Noise addition,  geometric transformation, randomization, condensation, and hybrid perturbation are a few of the perturbation approaches~\cite{zhong2012mu,chamikaraprocal}.  As data perturbation methods do not change the original input data formats, they may concede some privacy leak~\cite{machanavajjhala2015designing}. A privacy model defines the constraints on the level of privacy of a particular perturbation mechanism~\cite{machanavajjhala2015designing};  $k-anonymity$, $l-diversity$, $(\alpha, k)-anonymity$, $t-closeness$ and differential privacy (DP) are some of such privacy models~\cite{chamikaraprocal}.  DP was developed to provide a better level of privacy guarantee compared to previous privacy models that are vulnerable to different privacy attacks ~\cite{dwork2009differential,9000905}. Laplace mechanism, Gaussian mechanism ~\cite{chanyaswad2018mvg}, geometric mechanism, randomized response ~\cite{qin2016heavy}, and staircase mechanism ~\cite{kairouz2014extremal} are a few of the fundamental mechanisms used to achieve DP. There are many practical examples where these fundamental mechanisms have been used to build differentially private algorithms/methods. LDPMiner ~\cite{qin2016heavy}, PINQ~\cite{mcsherry2009privacy}, RAPPOR~\cite{erlingsson2014rappor}, and Deep Learning with DP~\cite{abadi2016deep} are a few examples of such practical applications of DP.

\section{Foundations of Differential Privacy and Eigenface recognition}
\label{fndatins}
In this section, we describe the background of the techniques used in the proposed solution. PEEP conducts privacy-preserving face recognition utilizing the concepts of differential privacy and eigenface recognition.

\subsection{Differential Privacy (DP)}
DP is a privacy model that is known to render maximum privacy by minimizing the chance of individual record identification~\cite{kairouz2014extremal}. In principle, DP defines the bounds to how much information can be revealed to a third party/adversary about someone's data being present in a particular database. Conventionally $\varepsilon$ (epsilon) is used to denote the level of privacy rendered by a randomized privacy-preserving algorithm ($\mathcal{M}$) over a particular database ($\mathcal{D}$); $\varepsilon$ is called the privacy budget that provides an insight into the privacy loss of a DP algorithm. The higher the value of $\varepsilon$, the higher the privacy loss. 

Let us take two adjacent datasets of $\mathcal{D}$, $x$ and $y$, where $y$ differs from $x$ only by (plus or minus) one person.  Then $\mathcal{M}$ satisfies ($\varepsilon$)-DP if Equation \eqref{dpeq} holds. Assume,  datasets $x$ and $y$ as being collections of records from a universe $\mathcal{X}$ and $\mathbb{N}$ denotes the set of all non-negative integers including zero.

\begin{mydef}
A randomized algorithm $\mathcal{M}$ with domain $\mathcal{N}^{|\mathcal{X}|}$ and
range $R$: is $\varepsilon$-differentially private  if for every adjacent $x$, $y$ $\in$ $\mathcal{N}^{|\mathcal{X}|}$
 and for any subset $\mathcal{S} \subseteq \mathcal{R}$
 \label{difpriv}
\end{mydef}
\begin{equation}
\mathcal{P}r[\mathcal{(M}(x) \in \mathcal{S})] \leq \exp(\varepsilon)~\mathcal{P}r[\mathcal{(M}(y) \in \mathcal{S})] 
\label{dpeq}
\end{equation}

\subsection{Global vs. Local Differential Privacy}

Global differential privacy (GDP) and local differential privacy (LDP) are the two main approaches to DP.  In the GDP setting, there is a trusted curator who applies carefully calibrated random noise to the real values returned for a particular query. The GDP setting is also called the trusted curator model~\cite{chan2012differentially}. Laplace mechanism and Gaussian mechanism~\cite{dwork2014algorithmic} are two of the most frequently used noise generation methods in GDP~\cite{dwork2014algorithmic}.  A randomized algorithm, $\mathcal{M}$ provides $\varepsilon$-GDP if Equation \eqref{dpeq} holds.  LDP randomizes data before the curator can access them, without the need of a trusted curator. LDP  is also called the untrusted curator model ~\cite{kairouz2014extremal}.  LDP can also be used by a trusted party to randomize all records in a database at once. LDP algorithms may often produce too noisy data, as noise is applied to achieve individual record privacy. LDP is considered to be a strong and rigorous notion of privacy that provides plausible deniability and deemed to be a state-of-the-art approach for privacy-preserving data collection and distribution. A randomized algorithm $\mathcal{A}$ provides $\varepsilon$-LDP if  Equation \eqref{ldpeq} holds ~\cite{erlingsson2014rappor}.

\begin{mydef}
A randomized algorithm $\mathcal{A}$  satisfies $\varepsilon$-LDP if for all pairs of users' inputs $v_1$ and $v_2$ and for all $\mathcal{Q} \subseteq Range(\mathcal{A})$, and for ($\varepsilon \geq 0$)  Equation \eqref{ldpeq} holds. $Range(\mathcal{A})$ is the set of all possible outputs of the randomized algorithm $\mathcal{A}$.
\end{mydef}

\begin{equation}
\mathcal{P}r[\mathcal{A}(v_1) \in \mathcal{Q}] \leq \exp(\varepsilon)~Pr[\mathcal{A}(v_2) \in \mathcal{Q}]
\label{ldpeq}
\end{equation}

\subsection{Sensitivity}

Sensitivity is defined as the maximum influence that a single individual can have on the result of a numeric query. Consider a function $f$, the sensitivity  ($\Delta f$) of $f$ can be given as in Equation \eqref{seneq} where x and y are two neighboring databases (or in LDP, adjacent records) and $\lVert . \rVert_1$ represents the $L1$ norm of a vector~\cite{wang2016using}. 

\begin{equation}
\Delta f=max\{\lVert f(x)-f(y) \rVert_1\}
\label{seneq}
\end{equation}

\subsection{Laplace Mechanism}

The Laplace mechanism is considered to be one of the most generic approaches to achieve DP~\cite{dwork2014algorithmic}. 
Laplace noise can be added to a function output ($\mathcal{F}(\mathcal{D})$) as given in Equation \ref{diffeq2} to produce a differentially private output. $\Delta f$ denotes the sensitivity of the function $f$. In local differentially private setting, the scale of the Laplacian noise is equal to $\Delta f/\varepsilon$, and the position is the current input value ($\mathcal{F}(\mathcal{D})$).

\begin{equation}
    \mathcal{PF}(\mathcal{D})= \mathcal{F}(\mathcal{D})+Lap(\frac{\Delta f}{\varepsilon})
\label{diffeq}
\end{equation}

\begin{equation}
    \mathcal{PF}(\mathcal{D})= \frac{\varepsilon}{2\Delta f}~e^{-\frac{|x-\mathcal{F}(\mathcal{D})|\varepsilon}{\Delta f}}
\label{diffeq2}
\end{equation}

\subsection{Eigenfaces and Eigenface recognition}
The process of face recognition involves data classification where input data are images, and output classes are persons' names. A face recognition algorithm needs to be first trained with an existing database of faces. The trained model will then be used to recognize a person's name using an image input. The training algorithm often needs various images to have high accuracy. When the model needs to be trained to recognize a large number of persons, the training algorithm also needs a large number of training images. Image data are often large, and the higher the number of faces to be trained, the slower the algorithm. However, facial recognition systems need high efficiency, as many of them are employed in real-time systems such as citizen surveillance~\cite{zhang1997face}. When an artificial neural network (ANN) is used for face recognition, the input images need to be flattened into 1-d vectors. An image with the dimensions $m\times n$ will result in an $mn\times 1$ vector. High-resolution images will result in extremely long 1-d vectors, which leads to slow training and testing of the corresponding ANN. Dimensionality reduction methods can be used to avoid such complexities, and allow face recognition to concentrate on the essential features, and to ignore the noise in the input images. In dimensionality reduction, the points are projected onto a higher-dimensional line, which is named as a hyperplane. Principal component analysis (PCA) is a dimensionality reduction technique that represents a hyperplane with maximum variance. This hyperplane can be determined using eigenvectors, which can be computed using the covariance matrix of input data ~\cite{zhang1997face}.

\begin{center}
    \scalebox{0.9}{
    \begin{minipage}{1.1\linewidth}
     \removelatexerror
      \begin{algorithm}[H]
    \caption{Generating Eigenfaces}
    \label{eigen_algo}
            \KwIn{
            \begin{tabular}{l c l} 
            $\{x^c_1,\dots, x^c_n\}               $ & $\gets $ & normalized and centered examples\\
              $nc$ & $\gets $ & expected number of PCA components\\
             \end{tabular}
             }
				
			\KwOut{
			\begin{tabular}{ l c l } 
				$\mathcal{EIMAT}$ & $ \gets $ &  matrix of eigenfaces \\
			\end{tabular}
			}
			\For{each $x^c_i$}{
			flatten $x^c_i$ to produce vector $t_i$}
			compute the mean face vector ($\mathcal{F}_m$), $\mathcal{F}_m=\frac{1}{n}\Sigma_{i=1}^{n}t_i$\;
			\For{each $x^c_i$}{
			$s_i=t_i - \mathcal{F}_m$\;
			}
			generate covariance matrix, $\mathcal{C}$,\newline
			$\mathcal{C}=\frac{1}{n}\Sigma_{i=1}^{n}s_i\times s_i^\mathcal{T}=\mathcal{AA}^\mathcal{T}$, where, $\mathcal{A}=[s_1s_2\dots s_n$]\;
			calculate the eigenvectors $e_i$ of $\mathcal{AA}^T$\newline
			since, $\mathcal{AA}^T$ can be extensive, derive $e_i$ from the eigenvectors $u_i$ of $\mathcal{A}^\mathcal{T}\mathcal{A}$, where, $e_i=\mathcal{A}u_i$\;
			compute the $n$ best eigenvectors $e_i$ such that, $\left\|e_{i}\right\|=1$\;
			return $nc$ eigenvectors which corresponds to the $nc$ largest eigenvalues

      \end{algorithm}
    \end{minipage}%
    }
     
  \end{center}

Algorithm \ref{eigen_algo} shows the steps for generating Eigenfaces. As shown in the algorithm, an eigenface~\cite{turk1991eigenfaces} utilizes PCA to represent a dimensionality-reduced version of an input image.   A particular eigenface considers a predefined number of the largest eigenvectors as the principal axes that we project our data on to, hence producing reduced dimensions~\cite{zhang1997face}. We can reduce the dimensions of an $m\times n$ image into a $k$ dimensional eigenface where $k$ is the largest $k$ eigenvectors. By doing this, we can consider only the most essential characteristics of an input image and increase the speed of a facial recognition algorithm while preserving high accuracy. Equation \ref{eigeq} provides the mathematical representation of an eigenface where $\mathcal{F}$ is a new face, $\mathcal{F}_m$ is the mean or the average face, $\mathcal{F}_i$ is an EigenFace, and $\alpha_i$ are scalar multipliers which we have to choose in order to create new faces.

\begin{equation}
\mathcal{F}=\mathcal{F}_m+\sum_{i=1}^{n}\alpha_i\mathcal{F}_i
\label{eigeq}
\end{equation}

\section{Our Approach: PEEP}
\label{ourapprch}
In this section, we discuss the steps employed in the proposed privacy-preserving face recognition approach (named as PEEP). We utilize DP to apply confidentiality to face recognition. PEEP applies randomization upon the eigenfaces to create privacy-preserving versions of input images.  We assume that any input device used to capture the facial images uses PEEP to apply randomization before sending the images to the storage devices/servers.

\begin{figure}[H]
	\centering
	\scalebox{0.48}{
	\includegraphics[width=1\textwidth, trim=0cm 0cm 0cm 0cm]{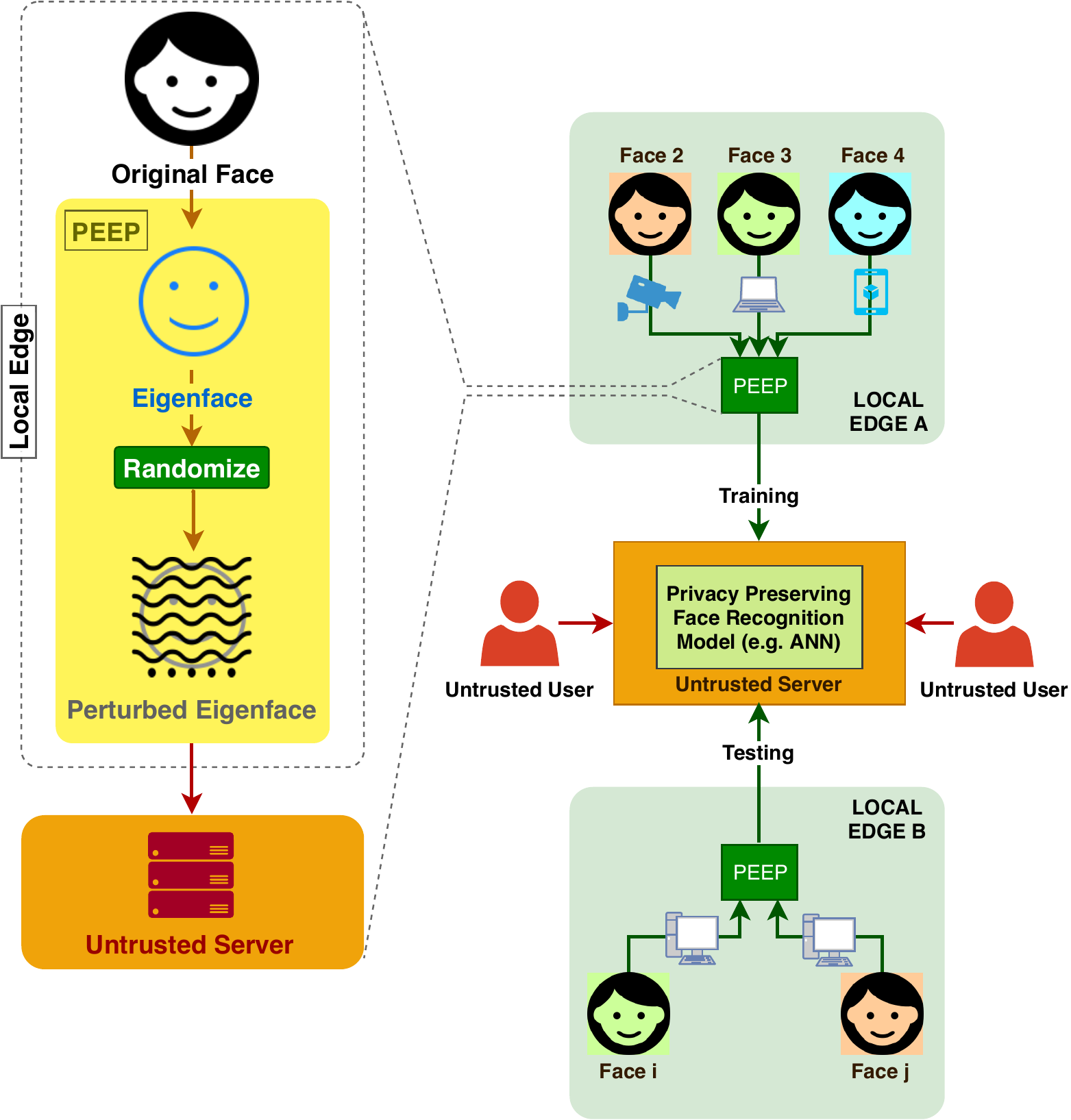}
	}
	\caption{Privacy-preserving face recognition using PEEP. The figure shows the placement of PEEP in a face recognition system. As shown, PEEP randomizes both training and testing images so that the untrusted third-party servers do not leak any private data to untrusted users. The callout figure in the left-hand side shows the basic flow of randomization inside PEEP, which applies Laplacian noise over eigenfaces.}
	\label{peepprivacymodel}
\end{figure}

As depicted by the callout box in Figure \ref{peepprivacymodel}, PEEP involves three primary steps to enforce privacy on face recognition.  They are, 1. accepting original face images, 2. generating eigenfaces, and 3. adding Laplacian noise to randomize the images.  In the proposed setting, the face recognition model (e.g. MLPClassifier) will be trained solely using randomized data.  In this setup, an untrusted server will hold only a privacy-preserving version of the face recognition model.

\subsection{Distributed eigenface generation}
When the number of input faces increases to a large number, it is important that the eigenface calculation (generation) can be distributed in order to maintain efficiency. Algorithm \ref{eigen_algo_dist} shows an incremental calculation approach of eigenfaces where a central computer (CC) in the local edge contributes to the calculation of eigenfaces in a distributed fashion.   As shown in step \ref{globmean} in Algorithm \ref{eigen_algo_dist}, the mean face vectors, $\mathcal{F}_m^i$ that are generated for each partition of input data are collected and merged (using Equation \ref{meancalc}) by the CC to generate the global mean face vector $\mathcal{F}_m^{glob}$. Similarly,  the CC generates the global covariance matrix, $\mathcal{C}^{glob}$ (refer step \ref{globcov} Algorithm \ref{eigen_algo_dist}) using the covariance matrices generated for each partition using Equation \ref{matcovupdate}. In this way, PEEP manages to maintain the efficiency of eigenface generation for extensive datasets.

\begin{equation}
\mathcal{F}_m^{glob}=
\begin{bmatrix}
\frac{m_{1}\times\overline{y_{11}}+m_{2}\times\overline{y_{12}}+\hdots+m_{k}\times\overline{y_{1k}}}{m_{1}+m_{2}+\hdots +m_{k}}\\
\frac{m_{1}\times\overline{y_{21}}+m_{2}\times\overline{y_{22}}+\hdots+m_{k}\times\overline{y_{2k}}}{m_{1}+m_{2}+\hdots +m_{k}}\\
\vdots\\
\frac{m_{1}\times\overline{y_{n1}}+m_{2}\times\overline{y_{n2}}+\hdots+m_{k}\times\overline{y_{nk}}}{m_{1}+m_{2}+\hdots +m_{k}}\\
\end{bmatrix}_{n\times 1}
\label{meancalc}
\end{equation}

In Equation \ref{meancalc}, $m_{i}$ refers to the number of eigenfaces in the $i^{th}$ partition, whereas $\overline{y_{ij}}$ refers to the mean of the $j^{th}$ index of the $i^{th}$ partition. To merge the covariance matrices, the pairwise covariance update formula introduced in \cite{bennett2009numerically} is adapted as shown in Equation \ref{matcovupdate}~\cite{chamikara2020privacy}. The pairwise covariance update formula for the two merged two column ($u$ and $v$) data partitions, $A$ and $B$, can be written as shown in Equation \ref{covupdate} where the merged dataset is denoted as $ X $.

\begin{equation}
\resizebox{0.8\textwidth}{!}{$\displaystyle
Cov(X)=\frac{\frac{C_A}{(m_A-1)}+\frac{C_B}{(m_B-1)}+(\mu_{u,A}-\mu_{u,B})(\mu_{v,A}-\mu_{v,B}).\frac{m_A.m_B}{m_X}}{(m_X-1)}
$}
\label{covupdate}
\end{equation}

Where, $\mu_{u,A}, \mu_{u,A}, \mu_{v,A}, \mu_{v,B}$ are means of $u$ and $v$ of the two data partitions $\ A $ and $\ B $, respectively.   $C_A$ and $C_B$ are the co-moments of the two data partitions $A$ and $B$ where the co-moment of a two column ($u$ and $v$) dataset $D$ is represented as,

\begin{equation}
C_D=\sum_{(u,v)\in D} (u-\mu_u)(v-\mu_v)
\end{equation}

Therefore, the variance-covariance matrix update formula of the two data partitions $\ D_g $ and $\ D_i $ can be written as shown in Equation \ref{matcovupdate},

\begin{equation}
\resizebox{0.9\textwidth}{!}{$\displaystyle
\mathcal{C}^{glob}=\frac{\frac{\mathcal{C}^{glob}}{(m_{D_g}-1)}+\frac{\mathcal{C}_i}{(m_{D_i}-1)}+(\mu_{D_g}(MI_g)-\mu_{D_i}(MI_g))(\mu_{D_g}(MI_i)-\mu_{D_i}(MI_i)).\frac{m_{D_g}.m_{D_i}}{m_{D_{new}}}}{(m_{D_{new}}-1)}
$}
\label{matcovupdate}
\end{equation}

In Equation \ref{matcovupdate}, assume that $\mathcal{C}^{glob}$ and $\mathcal{C}_i$  are the covariance matrices returned for the data partitions $\ D_g $ and $\ D_i $ respectively, where $D_g$ represents the global partition (concatenation of all the former partition), whereas $D_i$ represents the new partition introduced to the calculation. $\ D_{new} $ is the merged dataset of the the data partitions, $\ D_g $ and $\ D_i $. $\mu_{D_g} $ and $\mu_{D_i} $ are mean vectors of $\ D_g $ and $\ D_i $ respectively. $m_D$ represents the number of eigenfaces in the corresponding dataset.  Equation \ref{matcovupdate} will be iteratively calculated for all the data partitions to generate the final value of $\ D_g $. $\mathcal{C}^{glob}$ is initialized with the first partition, and $D_i$ will start from the second partition and, 

\begin{equation}
MI_i=
\begin{bmatrix}
[1]_n\\
[2]_n\\
[3]_n\\
\vdots\\
[n]_n\\
\end{bmatrix}_{n\times n}
\end{equation}

We can also run Algorithm \ref{eigen_algo_dist} in distributed computing nodes (DCN) within the local edge to conduct efficient eigenface generation. In such a setting, DCNs will communicate with a central computer (in the local edge) to generate the global mean face ($\mathcal{F}_m^{glob}$) and the global covariance matrix ($\mathcal{C}^{glob}$). In this way, an agency can deal with a large number of input faces by maintaining a feasible number of DCNs.

\begin{center}
    \scalebox{0.9}{
    \begin{minipage}{1.1\linewidth}
     \removelatexerror
      \begin{algorithm}[H]
    \caption{Incremental calculation of Eigenfaces using data partitions}
    \label{eigen_algo_dist}
            \KwIn{
            \begin{tabular}{l c l} 
            $\{x^{pk}_1,\dots, x^{pk}_n\}               $ & $\gets $ & normalized and centered example partition, $pk$\\
              $nc$ & $\gets $ & expected number of PCA components\\
             \end{tabular}
             }
				
			\KwOut{
			\begin{tabular}{ l c l } 
				$\mathcal{EIMAT}$ & $ \gets $ &  matrix of eigenfaces \\
			\end{tabular}
			}
			\For{each $x^{pk}_i$}{
			flatten $x^{pk}_i$ to produce vector $t_i$}
			compute the mean face vector ($\mathcal{F}_m^i$), $\mathcal{F}_m^i=\frac{1}{n}\Sigma_{i=1}^{n}t_i$\;
			collect $\mathcal{F}_m^i$ at a central computer (CC) in the local edge \;
			receive global mean face vector, $\mathcal{F}_m^{glob}$ from the CC\;\label{globmean}
			\For{each $x^c_i$}{
			$s_i=t_i - \mathcal{F}_m^{glob}$\;
			}
			generate covariance matrix, $\mathcal{C}_i$,\newline
			$\mathcal{C}_i=\frac{1}{n}\Sigma_{i=1}^{n}s_i\times s_i^\mathcal{T}=\mathcal{A}_i\mathcal{A}_i^\mathcal{T}$, where, $\mathcal{A}_i=[s_1s_2\dots s_n$]\;
			collect $\mathcal{C}_i$ at the CC\;
			receive global covariance matrix, $\mathcal{C}^{glob}$ from the CC\; \label{globcov} 
			
			calculate the eigenvectors $e_i$ of $\mathcal{AA}^T$, where $\mathcal{C}^{glob}$ = $\mathcal{AA}^T$\newline
			since, $\mathcal{AA}^T$ can be extensive, derive $e_i$ from the eigenvectors $u_i$ of $\mathcal{A}^\mathcal{T}\mathcal{A}$, where, $e_i=\mathcal{A}u_i$\;
			compute the $n$ best eigenvectors $e_i$ such that, $\left\|e_{i}\right\|=1$\;
			return $nc$ eigenvectors which corresponds to the $nc$ largest eigenvalues

      \end{algorithm}
    \end{minipage}%
    }
     
  \end{center}

\subsection{Generation of the principal components}
After accepting the image inputs, PEEP normalizes the images to match a predefined resolution (which is accepted by PEEP as an input). We consider a default resolution normalization of $47\times62$. However, based on the input image sizes and the computational power of the edge devices, the users can increase or decrease the values of $irw$ and $irh$ suitably. Following the steps of Algorithm \ref{eigen_algo}, PEEP calculates the principal components by considering the eigenvectors using the corresponding covariance matrix. The largest $nc$ (the number of principal components) number of eigenvectors are used to create a particular eigenface ($nc$ is taken as input). The higher the $nc$, the higher the representation of input features, the lower the efficiency. It is important to select a suitable number for $nc$ that can provide high accuracy and high efficiency simultaneously. A reliable number for $nc$ can be determined by investigating the change in the trained model's accuracy.

\subsection{Declaring the sensitivity before noise addition}

PEEP scales the indices of the identified PCA vectors within the interval [0,1] as the next step after generating the eigenfaces. In LDP, the sensitivity is the maximum difference between two adjacent records. In PEEP, the inputs are images, and each image is dimensionality reduced to form a vector by using PCA (PCA\_vectors). As PEEP adds noise to these vectors (PCA\_vectors), the sensitivity of PEEP is the maximum difference between two such PCA\_vectors which can be denoted by Equation \ref{seneq_new}, where $\mathcal{FSV}^j$ represents a flattened image vector scaled within the interval [0,1], $\mathcal{FSV}^{j+1}$ is adjacent to $\mathcal{FSV}^j$. Since PEEP examines the Cartesian system, we can consider the maximum Euclidean distance for the sensitivity, which is equal to a maximum of $\sqrt{nc}$ where $nc$ is the number of principal components. As the normalized PCA\_vectors are bounded by 0 and 1, a sensitivity much greater than 1 would entail a substantial level of noise, which can reduce the utility drastically as we use LDP for the noise application mechanism. Hence, we select the sensitivity to be the maximum difference between two indices, which is equal to 1.  Now the scale of the Laplacian noise will be equal to $1/\varepsilon$. As future work, we are conducting further algebraic analysis of sensitivity to improve the precision and flexibility of the Laplace mechanism in the proposed approach of face recognition. After defining the position and scale parameters, PEEP adds Laplacian noise to each index of PCA\_vectors. We take the position of the noise to be the index values and the scale of the noise to be $1/\varepsilon$.  To generate the private versions of images ($\mathcal{PI}$), we can perturb each index according to Equation \ref{diffeq2_new}, where $\mathcal{FSV}_i$ represents an index of the flattened image vectors scaled within the interval [0,1].

\begin{equation}
\Delta f=max\{\lVert \mathcal{FSV}^j-\mathcal{FSV}^{(j+1)} \rVert_1\}
\label{seneq_new}
\end{equation}

\subsection{Introducing Laplacian noise}
After defining the position and scale parameters, PEEP adds Laplacian noise to each index of PCA\_vectors. We take the position of the noise to be the index values and the scale of the noise to be $1/\varepsilon$.  To generate the private versions of images ($\mathcal{PI}$), we perturb each index according to Equation \ref{diffeq2_new}, where $FSV_i$ represents an index of the flattened image vectors scaled between 0 and 1. The user can provide a suitable $\varepsilon$ value depending on the amount of privacy required and after considering the following guidelines. The higher the $\varepsilon$ value, the lower the privacy.  As a norm,  $0<\varepsilon\leq 9$ is considered as an acceptable level of privacy~\cite{abadi2016deep}. We follow the same standard and use an upper limit of 9 for $\varepsilon$.

\begin{equation}
    \mathcal{PI}= \frac{\varepsilon}{2\Delta f}~e^{-\frac{|x-\mathcal{FSV}_i|\varepsilon}{\Delta f}}
\label{diffeq2_new}
\end{equation}

\begin{center}
    \scalebox{0.8}{
    \begin{minipage}{1.1\linewidth}
     \removelatexerror
      \begin{algorithm}[H]
    \caption{Differentially private facial recognition: PEEP}
    \label{ranalgo}
            \KwIn{
            \begin{tabular}{l c l} 
            $\{x_1,\dots, x_n\}               $ & $\gets $ & examples\\
            $imthresh$ & $\gets$ & number of images per face\\
              $\varepsilon              $ & $\gets $ & privacy budget\\
              $irw$                & $\gets $ & pixel width (default = 47)\\
              $irh$ & $\gets $ & pixel height (default = 62)\\
              $nc$ & $\gets $ & number of PCA components\\
             \end{tabular}
             }
				
			\KwOut{
			\begin{tabular}{ l c l } 
				$\mathcal{DPFRS}$ & $ \gets $ &  privacy preserving \\
				& & facial recognition model 
			\end{tabular}
			}
			Find the minimum width of all image ($w_{min}$)\;
			Fine the minimum height of all image ($h_{min}$)\;
			\If{$irw<w_{min}\lor  irh<h_{min}$ }{ \label{stepnc}
			   $irw=w_{min}$\\
			   $irh=h_{min}$
			}
			normalize the example resolution to $irw\times irh$ \label{step1}\;
            \If{$nc>irw\lor  nc>irh$ }{
			   $nc=min(irw,irh)$
			}
			generate the flattened vectors ($v_i$) for each $x_i$\;
            generate the first $nc$ PCA components ($\mathcal{PCA}_i$) for each input, $v_i$, according to Algorithm \ref{eigen_algo}\; 
             
            scale all the indices of $v_i$ between $0$ and $1$\ to generate $sv_i$\label{genind}\;
            apply $\frac{\varepsilon}{2\Delta f}~e^{-\frac{|x-\mathcal{FSV}_i|\varepsilon}{\Delta F}}$ to each index of $sv_i$ with  $sensitivity~(\Delta f)=1$ \label{step7}\label{applyrand}\;
            feed $\{sv_1,\dots, sv_n\}$ and corresponding targets to the classification model\;
            train the classification model using the randomized data to produce a differentially private classification model ($\mathcal{DPFRS}$)\;
            release the $\mathcal{DPFRS}$\;  
   
      \end{algorithm}
    \end{minipage}%
    }
     
  \end{center}

\subsection{Algorithm for generating a differentially private  face recognition model}
Algorithm \ref{ranalgo} shows the steps of PEEP in conducting privacy-preserving face recognition model training.  As shown in the algorithm, $irw$ and $irh$ parameters are used to increase the resolution of the input images.  We use the input parameter, $imthresh$, to accept the number of images considered per single face (person). Since the main task of face recognition is image classification, each face represents a class. In order to produce good accuracy, a classification model should have a good image representation. Consequently, $imthresh$ is a valuable parameter that directly influences the accuracy, where a higher value of $imthresh$ will certainly contribute to higher accuracy due to the better representation of images between the classes (faces). Hence, $imthresh$ allows the algorithm to extract eigenfaces that provide a better representation of the input images resulting in better accuracy. Step \ref{stepnc} makes sure that the number of PCA components selected does not go beyond the allowed threshold.

\subsection{Privacy preserving face recognition using PEEP}

As shown in Figure \ref{peepprivacymodel}, each image input will be subjected to PEEP randomization before training or testing.  The Eigenface generation and randomization take place within the local edge bounds. We assume that all input devices communicate with the third party servers only through PEEP, and the face recognition database stores only the perturbed images.  Since the face recognition model (e.g. MLPClassifier) is trained only using perturbed images (perturbed eigenfaces), the trained model will not leak private information. Any untrusted access to the server will not allow any loss of valuable biometric data to malicious third parties. Since PEEP perturbs testing data, there is minimal privacy leak from testing data (testing image inputs) as well.

\subsection{Theoretical privacy guarantee of PEEP on trained classifier}
Although additional computations are carried out on the outcome of a differentially private algorithm, they do not weaken the privacy guarantee. The results of additional computations on $\varepsilon$-DP outcome will still be $\varepsilon$-DP. This property of DP is called the postprocessing invariance/robustness~\cite{bun2016concentrated}.   Since PEEP utilizes DP, PEEP also inherits postprocessing invariance.  The postprocessing invariance property guarantees that the trained model of perturbed data satisfies the same privacy imposed by PEEP. Therefore, the proposed method ensures that there is a minimal level of privacy leak from the third party untrusted servers. However, we further investigate the privacy strength of PEEP using empirical evidence under Section \ref{resdis}.

\subsection{Datasets}
\label{datadetails}
\begin{figure}[H]
	\centering
	\scalebox{0.48}{
	\includegraphics[width=1\textwidth, trim=0cm 0cm 0cm 0cm]{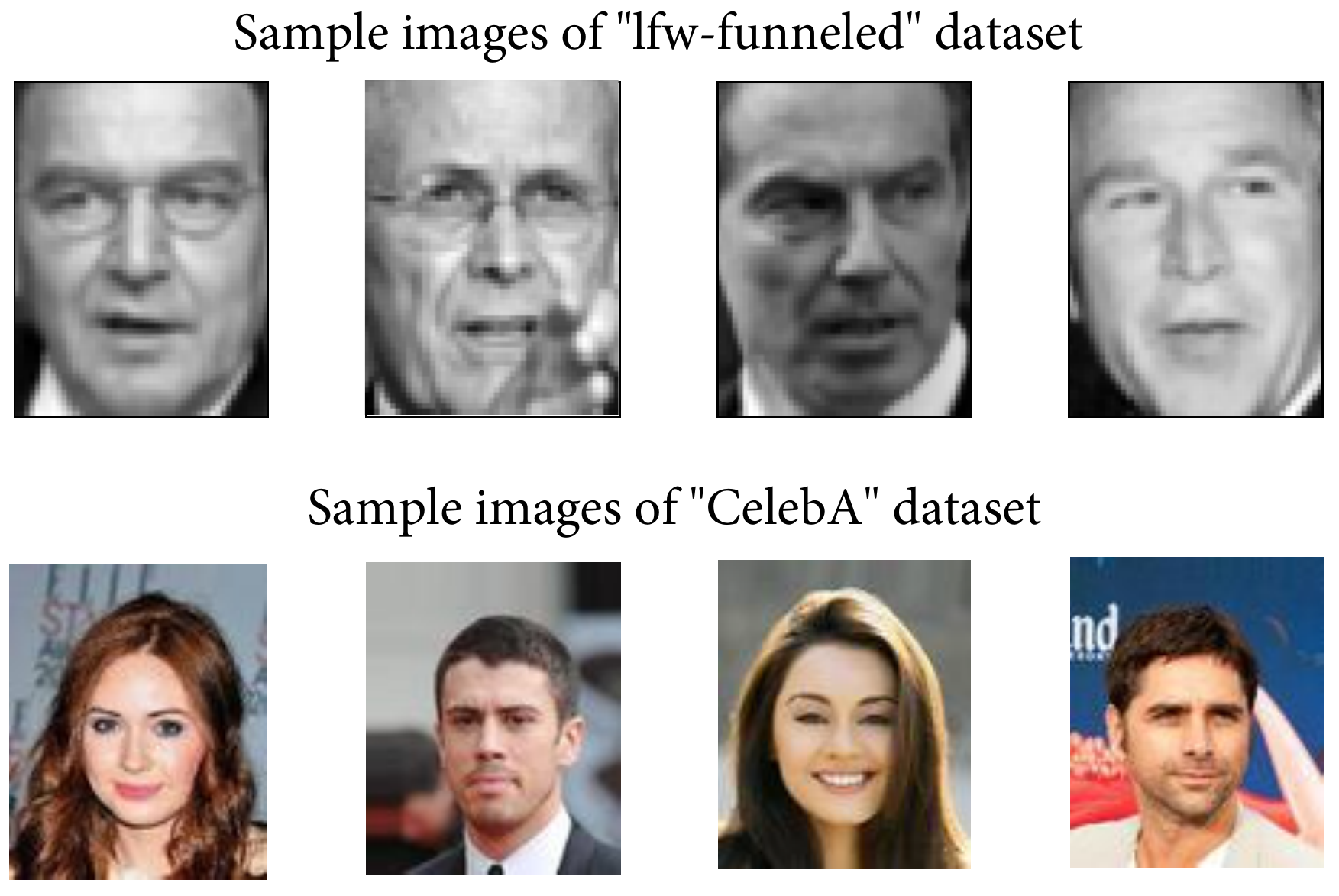}
	}
	\caption{Sample images of the two databases. The lfw-funneled dataset is composed of gray images whereas the CelebA dataset is composed of colored images. }
	\label{samplefigures}
\end{figure}

We used the open face image dataset and the large-scale CelebFaces Attributes (CelebA) dataset (see Figure \ref{samplefigures} for sample images) to test the performance of PEEP. Open face image dataset named lfw-funneled is available at the University of Massachusetts website named ``Labeled Faces in the Wild"\footnote{http://vis-www.cs.umass.edu/lfw/}.  The lfw-funneled dataset has 13,233 gray images. We limit the minimum number of faces per person to 100, which limits the number of images to 1,140 with five classes;   ``Colin Powell",  ``Donald Rumsfeld",  ``George W Bush", ``Gerhard Schroeder",  and  ``Tony Blair"\footnote{The diversity of the classes of the dataset are as follows, ``Colin Powell": 236, ``Donald Rumsfeld": 121, ``George W Bush": 530, ``Gerhard Schroeder": 109, and ``Tony Blair": 144.}. Figure \ref{samplefigures} shows the appearance of 8 sample images that are available in the datasets used. We used 70\% of the input dataset for training and 30\% for testing. CelebA\footnote{http://mmlab.ie.cuhk.edu.hk/projects/CelebA.html} dataset has more than 200K celebrity images, each with 40 attribute annotations.  CelebA has 10,177 number of identities, 202,599 number of face images, 5 landmark locations, and 40 binary attributes annotations per image.

\subsection{Eigenfaces and Eigenface perturbation}

\begin{figure}[H]
	\centering
	\scalebox{0.48}{
	\includegraphics[width=1\textwidth, trim=0cm 0cm 0cm 0cm]{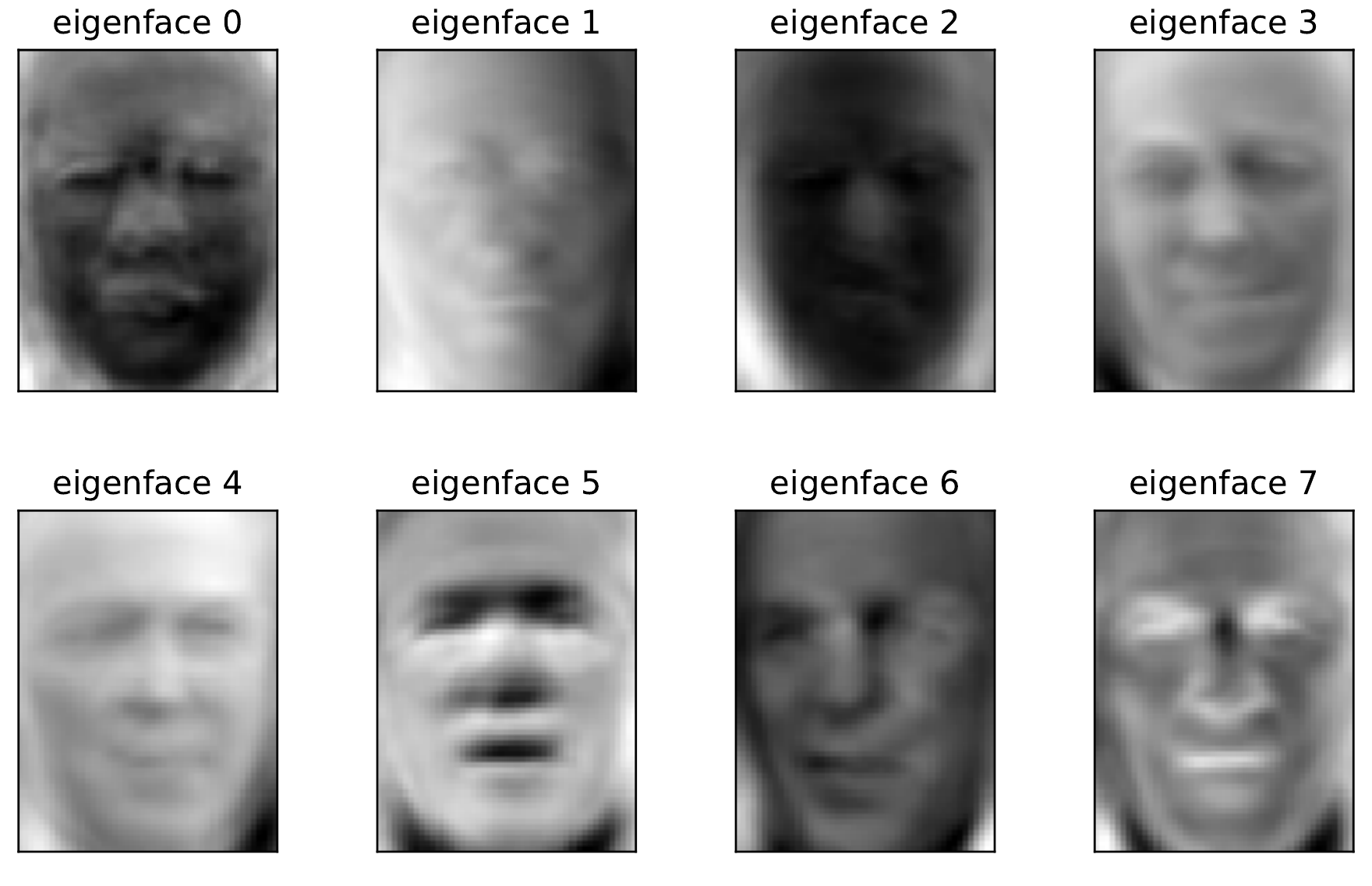}
	}
	\caption{Eigenfaces. The figure shows a collection of sample eigenfaces generated from the input face images. The eigenfaces show only the most essential features of the input images. }
	\label{eigenfaces}
\end{figure}

\begin{figure}[H]
	\centering
	\scalebox{0.48}{
	\includegraphics[width=1\textwidth, trim=0cm 0cm 0cm 0cm]{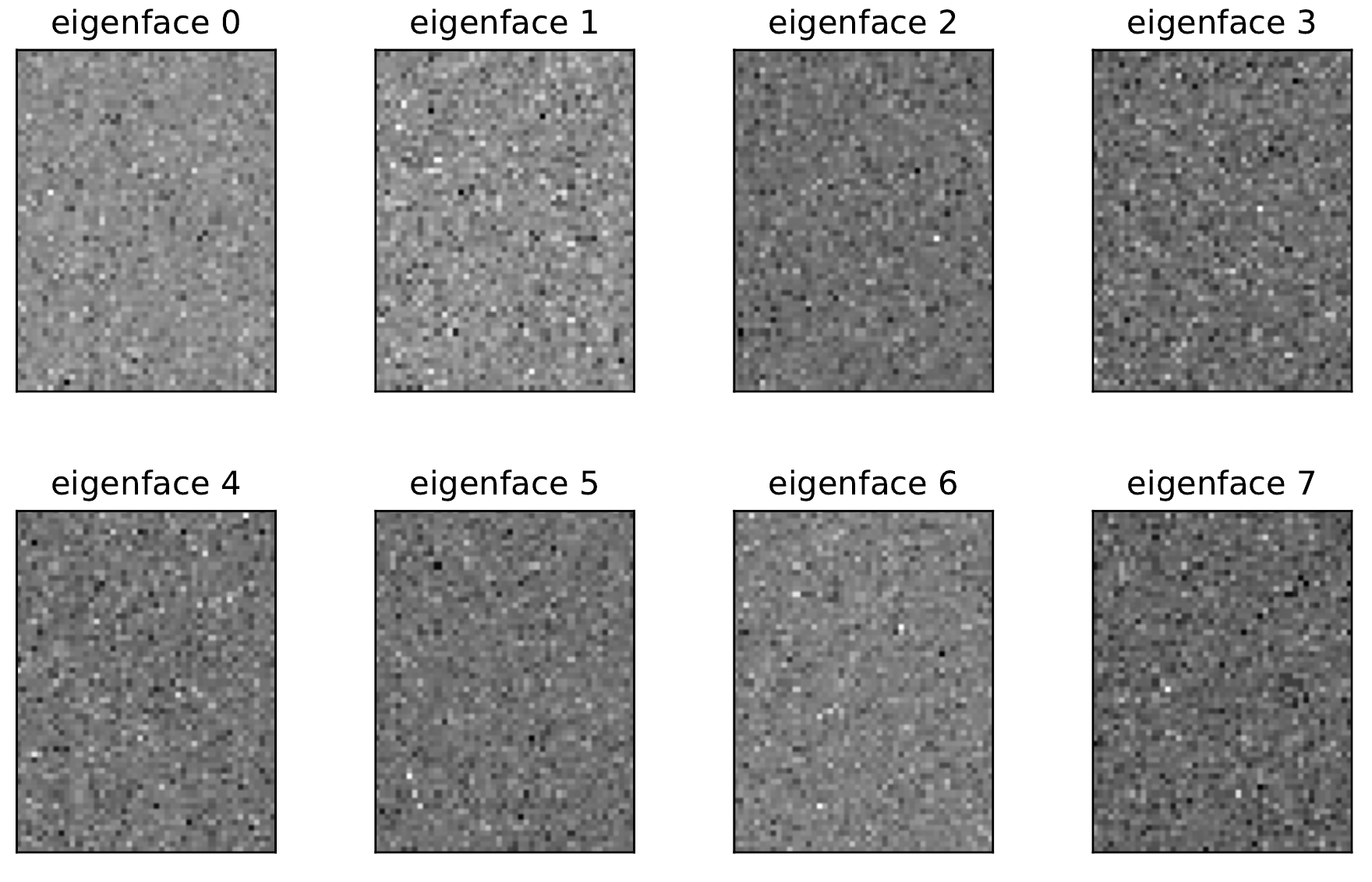}
	}
	\caption{Perturbed eigenfaces at $\varepsilon=4$. The randomized images appear to show no biometric features to the naked eye at $\varepsilon=4$.}
	\label{perteigen}
\end{figure}

\begin{figure}[H]
	\centering
	\scalebox{0.48}{
	\includegraphics[width=1\textwidth, trim=0cm 0cm 0cm 0cm]{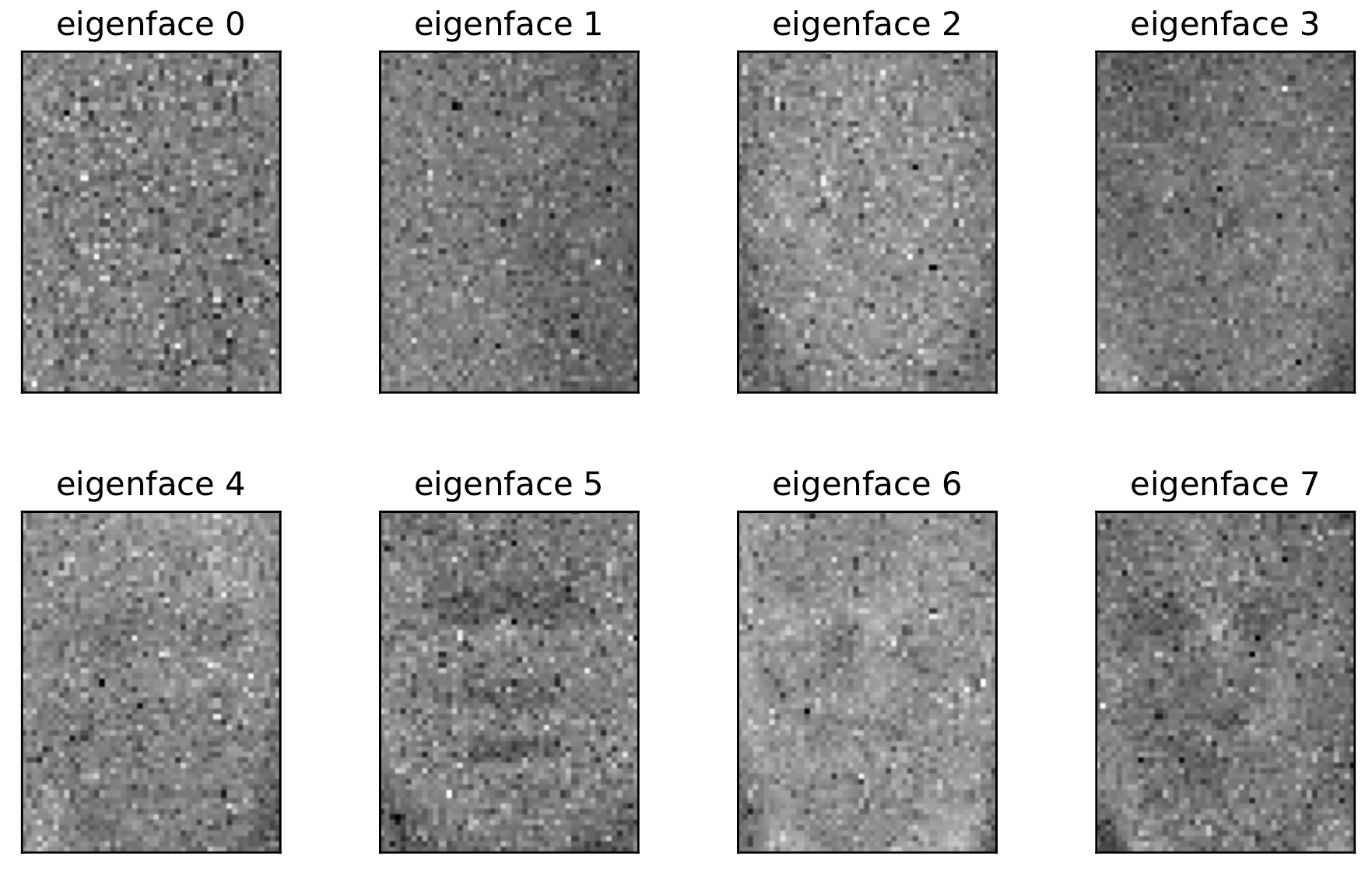}
	}
	\caption{Perturbed eigenfaces at $\varepsilon=100$. Here we try to demonstrate that even at an extreme case of the privacy budget (which is 100 and is not an acceptable value for $\varepsilon$, since $0<\varepsilon\leq 9$ is considered as the acceptable range for $\varepsilon$~\cite{abadi2016deep}), PEEP is capable of hiding a lot of biometric features from the eigenfaces.}
	\label{perteigen100}
\end{figure}

Figure \ref{eigenfaces} shows 8 sample eigenfaces before perturbation. As the figure shows, eigenfaces already hide some features of the original images due to the dimensionality reduction~\cite{aggarwal2004condensation}. However, eigenfaces alone would not provide enough privacy as they display the most important biometric features, and there are effective face reconstruction techniques~\cite{turk1991eigenfaces, pissarenko2002eigenface} for eigenfaces as demonstrated in Figure \ref{perteigen}, which shows the same set of eigenfaces (available in Figure \ref{eigenfaces}) after noise addition by PEEP with $\varepsilon=4$. As the figure shows, the naked eye cannot detect any biometric features from the perturbed eigenfaces. Even at an extreme case of a privacy budget $(\varepsilon=100)$, the perturbed eigenfaces show mild levels of facial features to the naked eyes, as shown in Figure \ref{perteigen100}. 

\section{Results and Discussion}
\label{resdis}

In this section, we discuss the experiments, experimental configurations, and their results. We used MLPClassifier to test the accuracy of face recognition with PEEP.  MLPClassifier is a multi-layer perceptron classifier available in the scikit learn\footnote{https://scikit-learn.org/stable/index.html} Python library.  We conducted all the experiments on a Windows 10 (Home 64-bit, Build 17134) computer with Intel (R) i5-6200U (6$^{th}$ generation) CPU (2 cores with 4 logical threads, 2.3 GHz with turbo up to 2.8 GHz) and 8192 MB RAM. Then we provide an efficiency comparison and a privacy comparison of  PEEP against two other privacy-preserving face recognition approaches developed by Zekeriya Erkin et al. (we abbreviate it as ZEYN for simplicity)~\cite{erkin2009privacy} and Ahman-Reza Sadehi et al. (we abbreviate it as  ANRA for simplicity) ~\cite{sadeghi2009efficient}. Both ZEYN and ANRA are cryptographic methods that use homomorphic encryption.

\subsection{Training the  MLPClassifier for perturbed eigenface recognition}
\label{mlpclassifier}
\begin{figure}[H]
	\centering
	\scalebox{0.4}{
	\includegraphics[width=1\textwidth, trim=0cm 0cm 0cm 0cm]{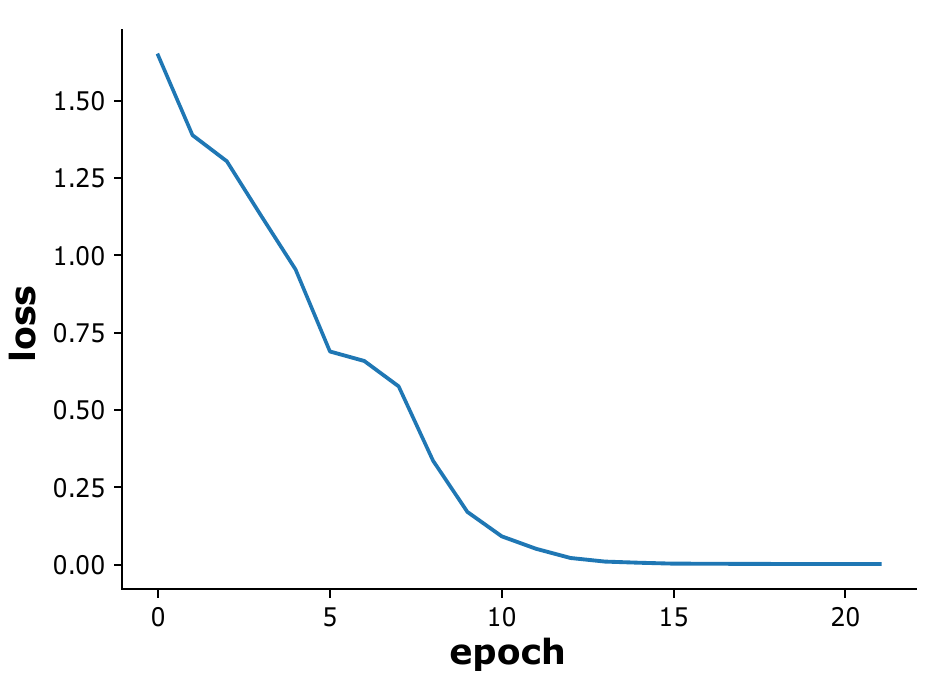}
	}
	\caption{Model loss when PEEP with $\varepsilon=4$. As shown in the figure, the MLPClassifier converges after around 14 epochs.}
	\label{modelloss4}
\end{figure}

We trained the MLPClassifier\footnote{Settings used for the MLP classifier; activation=`relu', batch\_size=100, early\_stopping =False, hidden\_layer\_sizes=(512, 1024, 2014, 1024, 512), max\_iter =200,  shuffle=True, and solver=`adam', alpha=0.0001, beta\_1=0.9,  beta\_2=0.999, epsilon=1e-08, learning\_rate=`constant', learning\_rate\_init=0.001, momentum=0.9, nesterovs\_momentum=True, power\_t=0.5, random\_state=None, tol=0.0001,   validation\_fraction=0.1, verbose=True, warm\_start=False.} under different levels of $\varepsilon$ ranging from 0.5 to 8 as plotted in Figure \ref{performancepeep}. Due to the heavy noise, the datasets with lower privacy budgets exhibited difficulty for training the MLPClassifier. However, we didn't conduct any parameter tuning to increase the performance of the MLPClassifier in order to make sure that we investigate the absolute impact of perturbation on the model. Figure \ref{modelloss4} shows the model loss of the training process of MLPClassifier when $\varepsilon=4$. As the figure shows, the model converges after around 14 epochs.

\subsection{Classification accuracy vs. privacy budget}

\begin{figure}[H]
	\centering
	\scalebox{0.5}{
	\includegraphics[width=1\textwidth, trim=0cm 0cm 0cm 0cm]{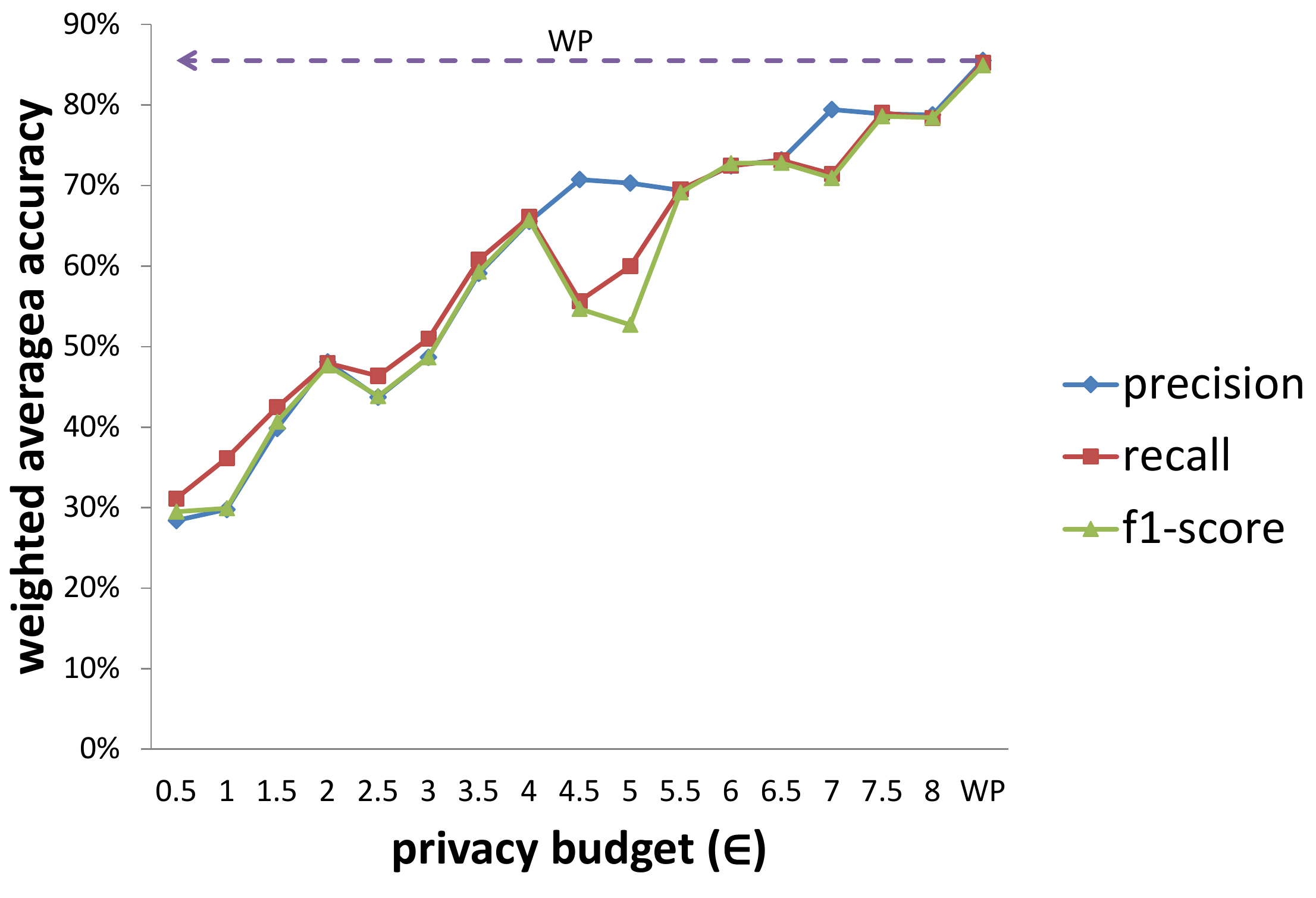}
	}
	\caption{Performance of face recognition with privacy introduced by PEEP. WP refers to the instance of classification model without privacy where no randomization is applied to the input images.}
	\label{performancepeep}
\end{figure}

We recorded the accuracy of the trained MLPClassifier in the means of the weighted average of precision, recall, and f$_1$-score against varying levels of privacy budget, and plotted the corresponding data as shown in Figure \ref{performancepeep}. As discussed in Section \ref{datadetails}, the class, ``George W Bush" showed a higher performance as there was a higher proportion of the input image instances related to that class. As shown in Figure \ref{performancepeep}, increasing the privacy budget increases accuracy, as higher privacy budgets impose less amount of randomization on the eigenfaces. We can see that PEEP produces reasonable accuracy for privacy budgets greater than 4 and less than or equal to 8, where $0<\varepsilon\leq 9$ is considered as an acceptable level of privacy~\cite{abadi2016deep}.

\begin{figure}[H]
	\centering
	\scalebox{0.48}{
	\includegraphics[width=1\textwidth, trim=0cm 0cm 0cm 0cm]{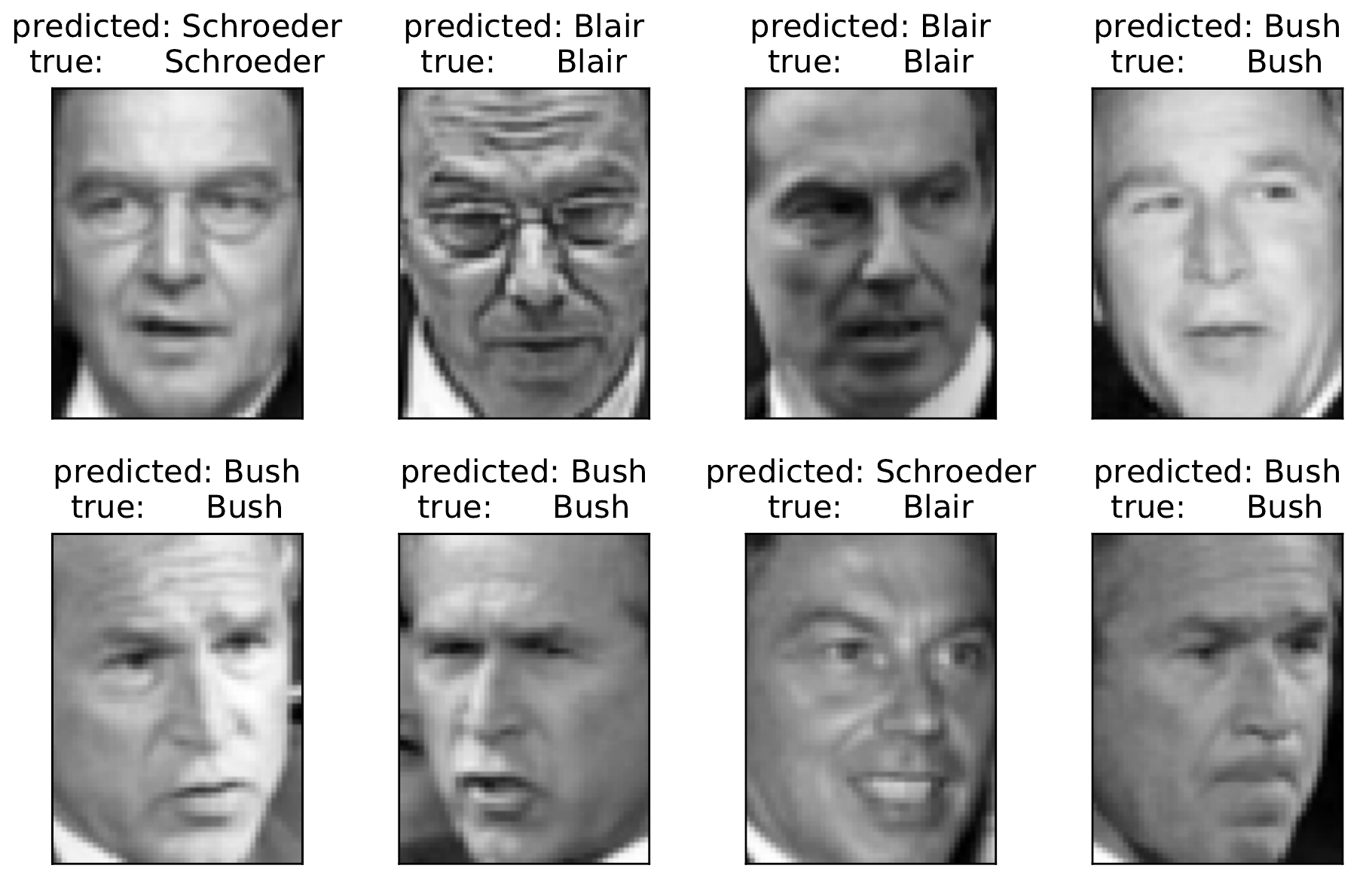}
	}
	\caption{Instance of the face recognition when the images are randomized using PEEP at $\varepsilon=4$ (the randomized images at $\varepsilon=4$ are shown in Figure \ref{perteigen}). The figure shows the predicted labels of the images against the original true labels. }
	\label{instface4}
\end{figure}

Figure \ref{instface4} shows the classification results of 8 random input images in the testing sample at $\varepsilon=4$. According to the figure, only in one case out of eight have been misclassified.  The parameters such as the minimum number of faces per each class, the size of the input dataset, and the hyperparameters of the MLPClassifier have a direct impact on accuracy. We can improve the accuracy of the MLPClassifier by changing the input parameters and conducting hyperparameter tuning. Moreover, the dataset has a  higher number of instances for the class ``George W Bush" compared to the other classes. A more balanced dataset would also provide better accuracy.  However, in this paper, we investigate only the absolute effect of the privacy parameters on the performance of the MLPClassifier.

\subsection{Effect of $imthresh$ on the performance of face recognition}
\begin{figure}[H]
	\centering
	\scalebox{0.5}{
	\includegraphics[width=1\textwidth, trim=0cm 0cm 0cm 0cm]{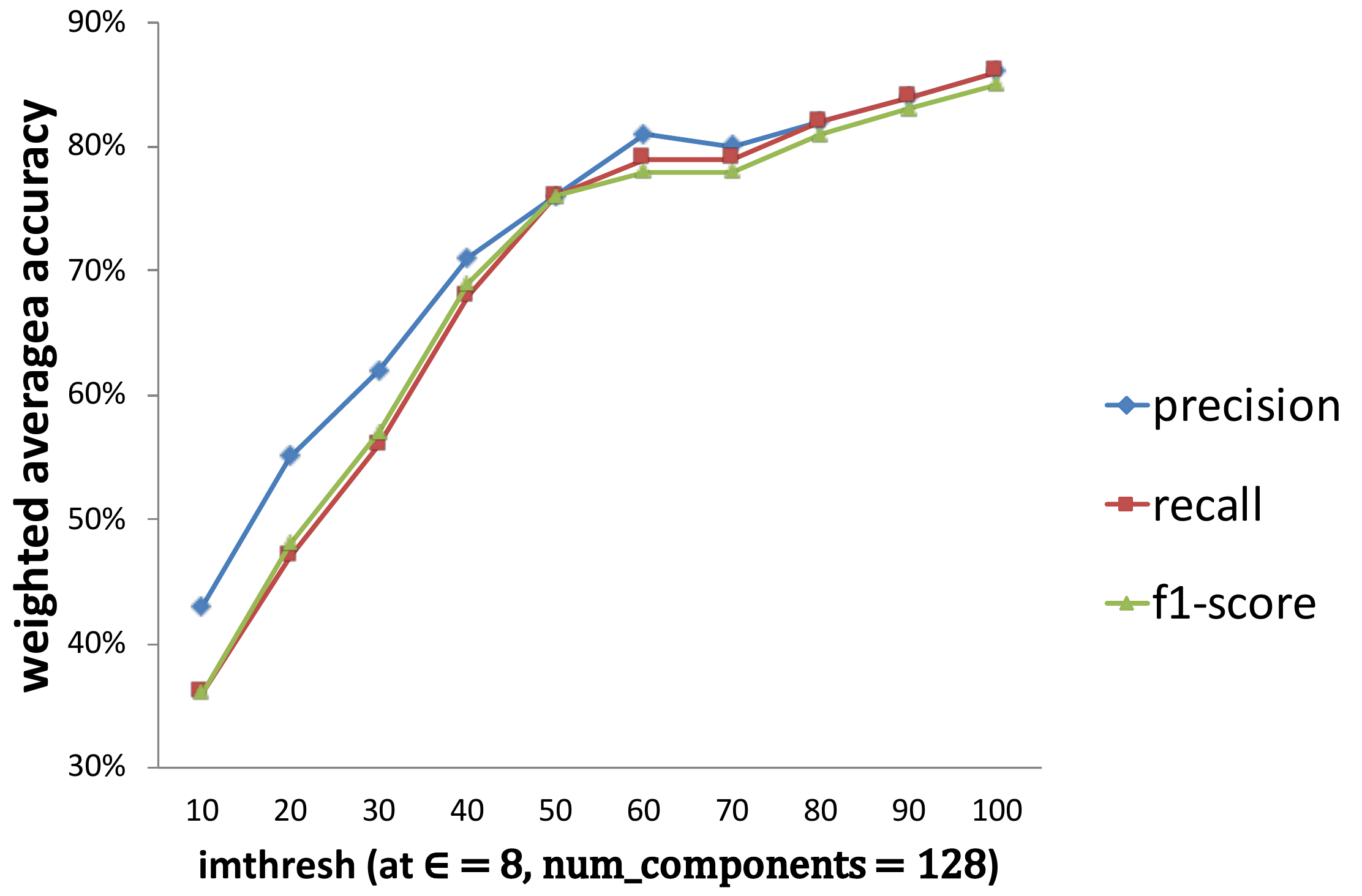}
	}
	\caption{Performance of face recognition Vs. $imthresh$.}
	\label{performthresh}
\end{figure}

In this section, we test the effect of $imthresh$ (the number of images per single face) on the performance of face recognition (refer to Figure \ref{performthresh}). During the experiment, we maintained an $\varepsilon$ value of 8 and the number of PCA components at 128.  As shown in the plots, the performance of classification improves with $imthresh$. This is a predicted observation as face recognition is a classification problem.  A higher value of $imthresh$ provides a higher representation for the corresponding face (class), generating higher accuracy. Hence, the proposed concept prefers a higher value for $imthresh$.  This feature encourages having the highest value possible for $imthresh$, in order to generate the highest accuracy possible.

\subsection{Effect of the number of PCA components on the performance of face recognition}

\begin{figure}[H]
	\centering
	\scalebox{0.5}{
	\includegraphics[width=1\textwidth, trim=0cm 0cm 0cm 0cm]{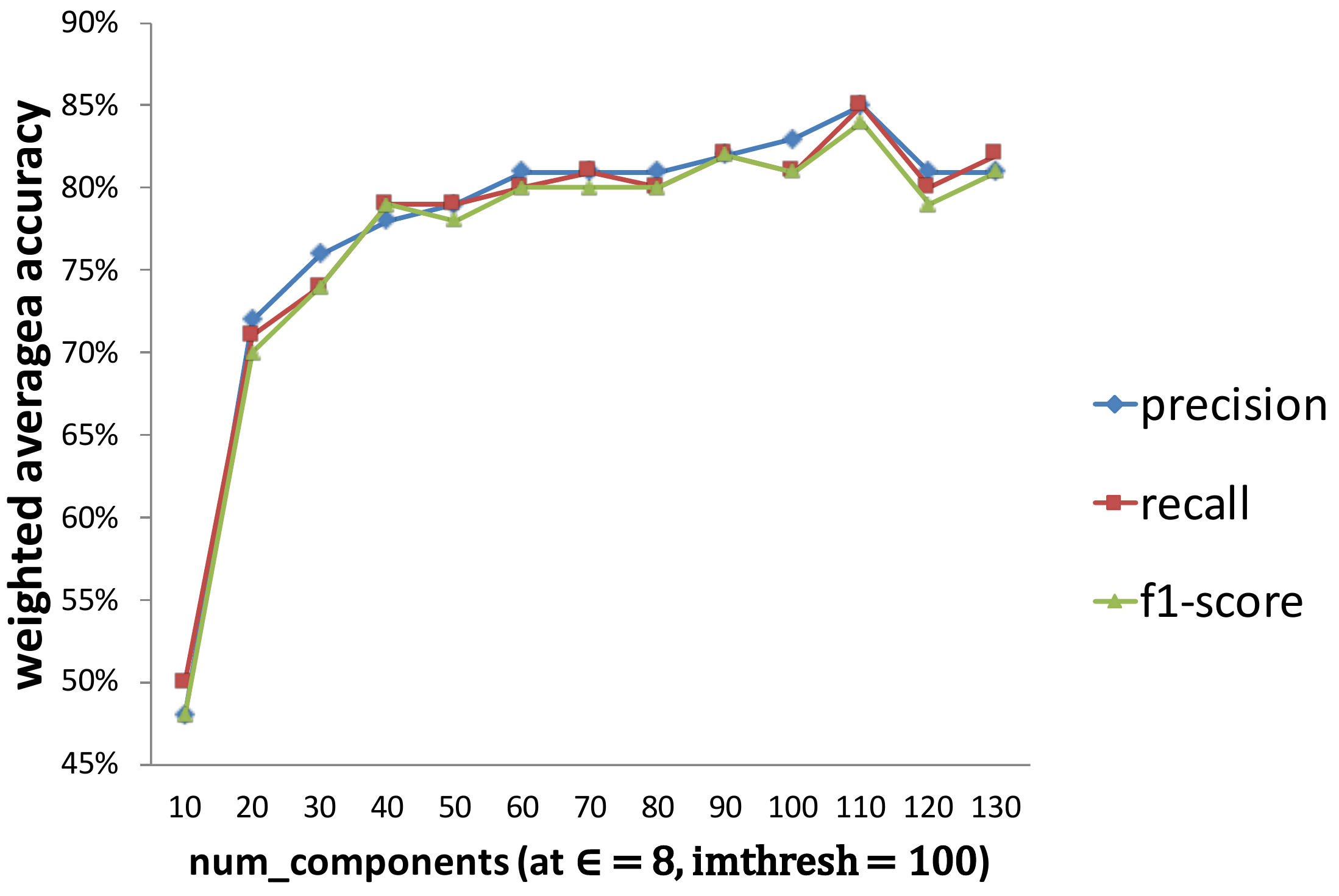}
	}
	\caption{Performance of face recognition Vs. the number of PCA components.}
	\label{performancenumcomponents}
\end{figure}

In this section, we investigate the effect of the number of PCA components on the performance of face recognition. During the experiment, we maintained an $\varepsilon$ value of 8, and $imthresh$ was maintained at 100. As shown by the plot (refer Figure \ref{performancenumcomponents}), there is an immediate increment of performance when the number of PCA components increased from 10 to 20. As the number of PCA components increase, there is a gradual increase in performance after 20 PCA components. This is due to the first 20 to 40 PCA components representing the most significant features of the input images. Although the effect of the number of PCA components after 40 is low, the improved performance suggests that it is better to have a higher number of PCA components to produce better performance.

\subsection{Face reconstruction attack setup}
\label{facereconsec}
It is essential that the randomized images cannot be used to reconstruct the original images that reveal the owners' identities. We prepared an experimental setup to investigate the robustness of PEEP against face reconstruction~\cite{turk1991eigenfaces,pissarenko2002eigenface} applied by adversaries on the randomized images. 

\begin{figure}[H]
	\centering
	\scalebox{0.34}{
	\includegraphics[width=1\textwidth, trim=0cm 0cm 0cm 0cm]{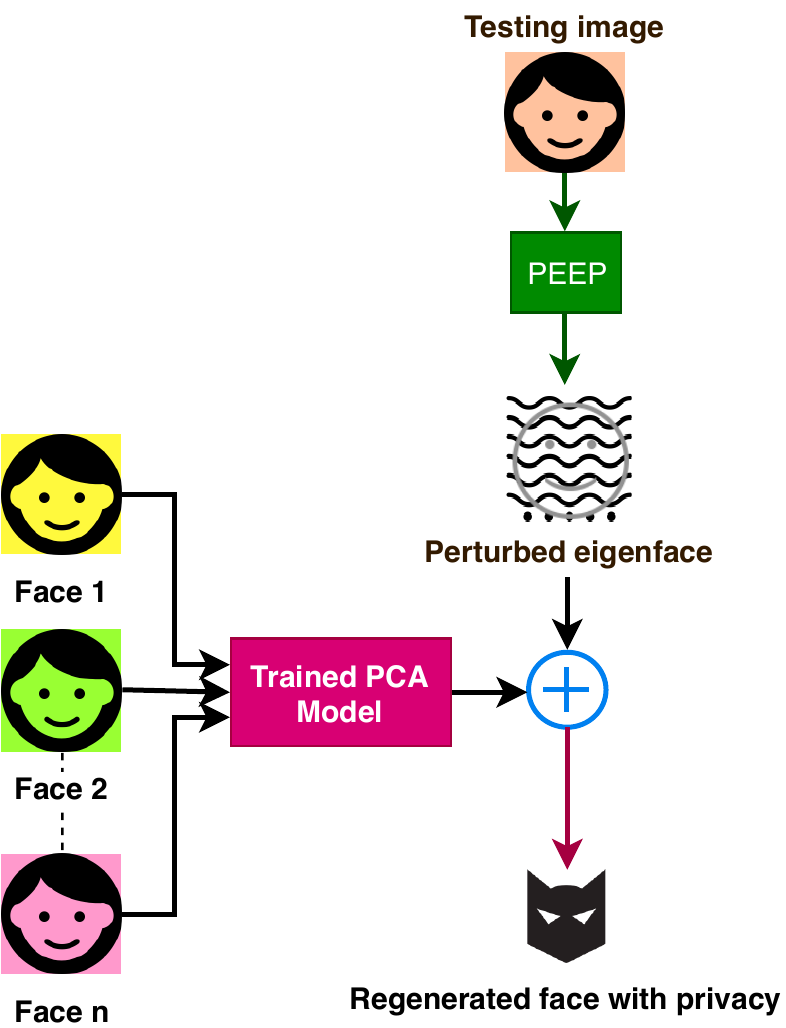}
	}
	\caption{Face reconstruction from perturbed eigenfaces. The figure shows the experimental setup used for the reconstruction of the original input face images using the perturbed eigenfaces.}
	\label{facereconstructattack}
\end{figure}

As shown in Figure \ref{facereconstructattack}, first, we create a PCAmodel (PCA: Principal Component Analysis)  using 2,000 training images (first 1,000 images of the CelebA database and the vertically flipped versions of them). The resolution of each image is $89\times 109$. The trained PCAmodel has the 2,000 eigenvectors of length 29,103 ($89\times 109\times 3$), and the mean vector (of 2,000 eigenvectors) of length 29,103. Next, the testing image (of size $89\times 109\times$) is read and flattened to form a vectorized form of the original image. The mean vector is then subtracted from it, and the resulting vector is randomized using PEEP to generate the privacy-preserving representation of the testing vector ($\mathcal{PV}$). Finally, we generate the eigenfaces ($\mathcal{F}_i$) and the average face by reshaping the eigenvectors ($\mathcal{FV}_i$) and mean vector available in the PCAmodel. Now we can reconstruct the original testing image from $\mathcal{PV}$ using Equation \ref{reconst1} where $n$ is the number of training images used for the PCAmodel, and $\mathcal{RI}$ is the recovered image.

\begin{equation}
\mathcal{RI}=\sum_{i=1}^{n} \mathcal{F}_i\times(\mathcal{PV}\bullet \mathcal{FV}_i) 
\label{reconst1}
\end{equation}

\subsection{Empirical privacy of PEEP}
\label{empriv}
\begin{figure}[H]
	\centering
	\scalebox{0.45}{
	\includegraphics[width=1\textwidth, trim=0cm 0cm 0cm 0cm]{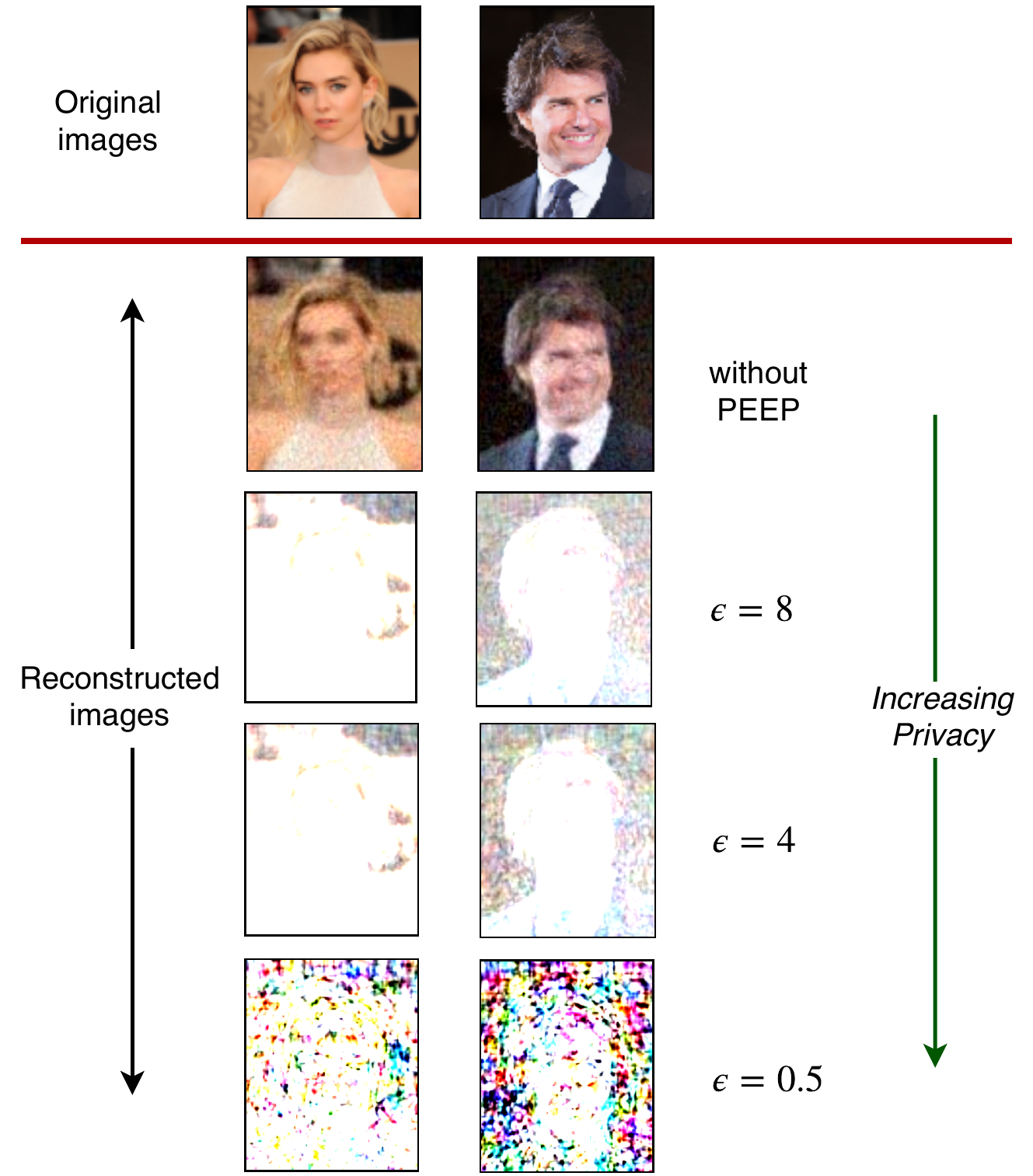}
	}
	\caption{Reconstructing images using the setup depicted in Figure \ref{facereconstructattack}. The first row shows original images.  The second row shows the reconstructed images using the eigenfaces of the first row images without privacy. The three remaining rows show the face reconstruction at the privacy levels of $\varepsilon$ equals to $8, 4$, and $0.5$, respectively.}
	\label{facereconstruct}
\end{figure}

Figure \ref{facereconstruct} shows the effectiveness of eigenface reconstruction attack (explained in Section \ref{facereconsec}) of a face image. The figure includes the results of the attack on two testing images.  Figure \ref{perteigen} provides the empirical evidence to the level of privacy rendered by PEEP in which the lower the $\varepsilon$, the higher the privacy. At $\varepsilon=0.5$, the attack is not successful in generating any underlying features of an image. At $\varepsilon=4$ and above, we can see that the reconstructed images have some features, but they are not detailed enough to identify the person shown in that image.

\subsection{Performance of PEEP against other approaches}

In this section, we discuss the privacy guarantee of PEEP and the comparable methods with regards to five privacy issues (TYIS 1, 2, 3, 4, and 5) in face recognition systems, as identified in Section \ref{intro}. The first six rows of Table \ref{privacyprotect} provide the summary of the evaluation, where a tick mark indicates effective addressing of a particular issue, while a cross mark shows failure. Partially addressed issues are denoted by a ``$\partial$" symbol. PEEP satisfies TYIS 1 and TYIS 4 by randomizing the input images (both training and testing) so that the randomized images do not provide any linkability to other sensitive data. Both ZEYN and ANRA are semi-honest mechanisms and need database owners to maintain the facial image databases. ZEYN and ANRA satisfy TYIS 1, if and only if the database owners are fully trusted, which can be challenging in a cloud setting, as untrusted third parties with malicious intent can access the cloud servers. As shown in Section \ref{empriv}, the randomized eigenfaces cannot be used to reconstruct original images. As the PEEP stores only randomized data in the servers, PEEP does not have to worry about the security of the cloud server. As a result, any data leak from the cloud server will not have an adverse effect on user privacy.  The scalability results of the three methods given in the last row of Table \ref{privacyprotect}  show that PEEP satisfies TYIS 2 by providing better scalability than ZEYN and ANRA. PEEP satisfies TYIS 3 because it uses no trusted party, whereas ZEYN and ANRA must have trusted database owners. PEEP provides some level of guarantee towards TYIS 5 by randomizing all the subsequent face image inputs related to the same person, which can come from the same device or different devices. Consequently, two input images related to the same person will have two different levels of randomization, leaving a low probability of linkability.

\begin{table}[H]
 
  \centering
  \caption{Performance of PEEP against other approaches}
  \resizebox{0.65\columnwidth}{!}{%
    \begin{tabular}{|l|l|c|c|c|}
    \hline
    \multirow{2}{*}{\begin{tabular}[c]{@{}l@{}}Qualitative\\ comparison\end{tabular}} &  \begin{tabular}[c]{@{}l@{}}Type of issue\\ (TYIS)\end{tabular} & ZEYN  & ANRA  & PEEP \\
    \cline{2-5}
    & \begin{tabular}[c]{@{}l@{}}1. biometric should \\not be  linkable to \\  other sensitive data \end{tabular}  & \textcolor[rgb]{ .329,  .51,  .208}{ \large $\partial$}  & \textcolor[rgb]{.329,  .51,  .208}{\large $\partial$} & \textcolor[rgb]{ .329,  .51,  .208}{\checkmark} \\
    \cline{2-5}
   &  \begin{tabular}[c]{@{}l@{}}2. scalable and \\ resource friendly\end{tabular} & \textcolor[rgb]{ 1,  0,  0}{\Large $\times$} & \textcolor[rgb]{ 1,  0,  0}{\Large $\times$} & \textcolor[rgb]{ .329,  .51,  .208}{\checkmark} \\
    \cline{2-5}
  &  \begin{tabular}[c]{@{}l@{}}3. biometrics should \\not be accessible\\ by a third-party\end{tabular} & \textcolor[rgb]{ 1,  0,  0}{\Large $\times$} & \textcolor[rgb]{ 1,  0,  0}{\Large $\times$} & \textcolor[rgb]{ .329,  .51,  .208}{\checkmark } \\
    \cline{2-5}
  &   \begin{tabular}[c]{@{}l@{}}4. biometrics of the\\ same person from\\ two applications\\ should not be\\linkable\end{tabular} & \textcolor[rgb]{ .329,  .51,  .208}{ \large $\partial$}  & \textcolor[rgb]{ .329,  .51,  .208}{ \large $\partial$}  & \textcolor[rgb]{ .329,  .51,  .208}{\checkmark } \\
    \cline{2-5}
 &   \begin{tabular}[c]{@{}l@{}}5. biometrics should\\ be revocable\end{tabular} & \textcolor[rgb]{ 1,  0,  0}{\Large $\times$} & \textcolor[rgb]{ 1,  0,  0}{\Large $\times$} & \textcolor[rgb]{ .329,  .51,  .208}{ \large $\partial$}  \\
    \hline
    \hline
  \multirow{2}{*}{\begin{tabular}[c]{@{}l@{}}Quantitative\\ comparison\end{tabular}} & \begin{tabular}[c]{@{}l@{}}Average time to \\ recognize one\\  image in seconds\\  when  the database\\ has 798 images\end{tabular} & \begin{tabular}[c]{@{}l@{}} $\sim$ 24  to 43\end{tabular} & \begin{tabular}[c]{@{}l@{}}$\sim$ 10 \\ \end{tabular} & 0.006 \\
    \hline
    \end{tabular}%
    }
    \newline\newline
    \textcolor[rgb]{ .329,  .51,  .208}{\checkmark } = fully satisfied, \textcolor[rgb]{ .329,  .51,  .208}{ \large $\partial$} = partially satisfied, \textcolor[rgb]{ 1,  0,  0}{\Large $\times$} = not satisfied
  \label{privacyprotect}%
  
\end{table}%

\subsection{Computational complexity}

\begin{figure}[H]
	\centering
	\scalebox{0.5}{
	\includegraphics[width=1\textwidth, trim=0cm 0cm 0cm 0cm]{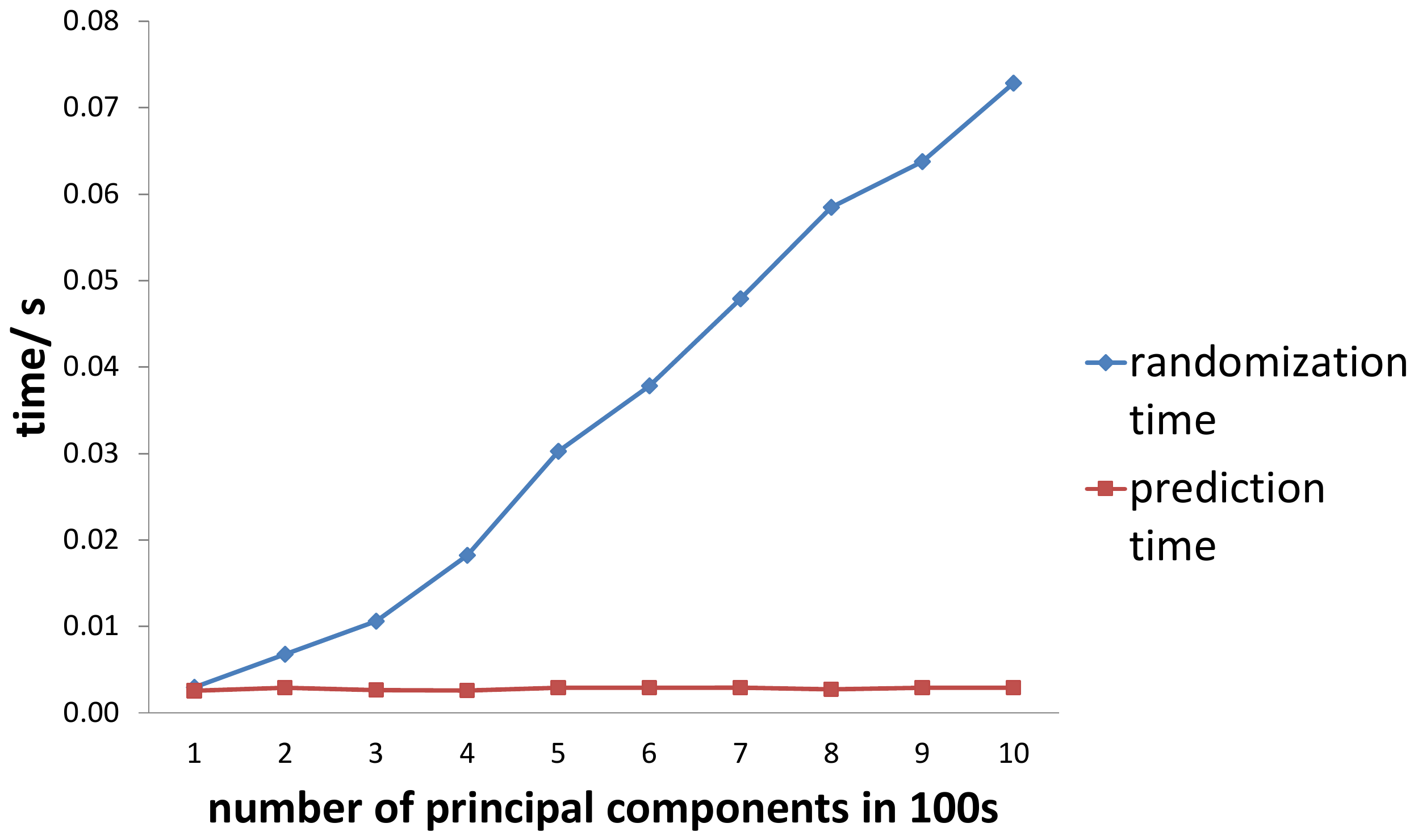}
	}
	\caption{The time consumption of PEEP to randomize and recognize one input image against the increasing number of principal components used for the eigenface generation. }
	\label{comcomplex}
\end{figure}

PEEP involves two independent segments (components) in recognizing a particular face image. Component 1 is the randomization process, and component 2 is the recognition process. The two components conduct independent operations; hence they need independent evaluations for computational complexity. Moreover, as PEEP does not need a secure communication channel, the complexity behind maintaining a secure channel does not affect the performance of PEEP.  For a particular instance of  PEEP (refer to Algorithm \ref{ranalgo}), step \ref{genind} to step \ref{applyrand} display linear complexity of $O(nc)$, where $nc$ is the number of principal components, and the image resolution (width in pixels, height in pixels) will remain constant during a particular instance of perturbation and recognition. When width in pixels=47, height in pixels=62, and the number of PCA components=128, PEEP takes around 0.004 seconds to randomize a single input image. Component 2 can be composed of any suitable classification model; in our case, we use the MLPClassifier (refer Section \ref{mlpclassifier}) as the facial recognition module that was trained using 798 images. Under the same input settings (width in pixels=47, height in pixels=62, and the number of PCA components=128), the trained model takes 0.002 seconds to recognize a facial image input. Since the prediction is always done on a converged model, the time taken for prediction will be constant and follow a complexity of $O(1)$. For randomization and prediction PEEP roughly consumes around 0.006 seconds under the given experimental settings. The runtime plots shown in Figure \ref{comcomplex} further validate the computational complexities evaluated above.  According to the last row of Table \ref{privacyprotect}, PEEP is considerably faster than comparative methods; PEEP provides a more effective and efficient approach towards the recognition of images against millions of faces in a privacy-preserving manner. In further examining the performance of PEEP for its efficiency, we investigated PE-MIU~\cite{terhorst2020pe}, and POR~\cite{ma2019lightweight}  (refer to Section \ref{relwork}), which are two recently developed approaches. PE-MIU consumes a complete MIU-verification time of 0.0072 seconds for a block size of 4 in a computer with an Intel(R) Core(TM) i7-7700 processor. POR consumes a testing time of around 0.011 seconds per one image in an Intel(R) Core(TM) i5-7200 CPU @2.50GHz and 8.00GB of RAM.  Hence, under the proposed experimental settings, a prediction time of 0.006 seconds consumed by PEEP can be considered as efficient and reliable.

\section{Conclusions}
\label{concls}
We proposed a novel mechanism named PEEP for privacy-preserving face recognition using data perturbation. PEEP utilizes the properties of differential privacy, which can provide a strong level of privacy to facial recognition technologies. PEEP does not need a trusted party and employs a local approach where randomization is applied before the images reach an untrusted server. PEEP forwards only randomized data, which requires no secure channel. PEEP is an efficient and lightweight approach that can be easily integrated into any resource-constrained device. As the training and testing/recognition of facial images done solely on the randomized data, PEEP does not incur any efficiency loss during the recognition of a face. The differentially private notions allow users to tweak the privacy parameters according to domain requirements.  All things considered, PEEP is a state of the art approach for privacy-preserving face recognition. 
 
Using the proposed approach with different biometric algorithms and areas like fingerprint and iris recognition will be looked at in the future, in particular with regards to effectiveness and sensitivity in different domains of inputs.

\end{document}